\documentclass[twocolumn,dvipsnames]{aastex63}

\graphicspath{{./}{figures/}}
\submitjournal{ApJ}
\accepted{2022 February 24}

\shorttitle{2693}
\shortauthors{Pilawa et al.}

\newcommand{\ngc}{NGC~2693}

\newcommand{\kms}{\ensuremath{}{\rm \, km~s^{-1}}}
\newcommand{\msun}{\ensuremath{M_{\odot}}}
\newcommand{\mbh}{\ensuremath{M_\mathrm{BH}}}
\newcommand{\ml}{\ensuremath{M^*/L}}
\newcommand{\vc}{\ensuremath{V_c}\ }

\newcommand{\GG}[1]{}
\newcommand{\tmaj}{\ensuremath{T_\text{maj}}}
\newcommand{\tmin}{\ensuremath{T_\text{min}}}

\usepackage{xcolor}

\newcommand{\clt}[1]{{\color{black} #1}}
\newcommand{\cpm}[1]{{\color{black} #1}}

\newcommand{\citeColored}[2]{{\hypersetup{citecolor=white}\citeauthor{#2}}}

\let\tablenum\relax
\usepackage{siunitx}
\usepackage{amsmath}

\begin{document}

\title{The MASSIVE Survey - XVII. A Triaxial Orbit-based Determination of the Black Hole Mass and Intrinsic Shape of Elliptical Galaxy NGC~2693}

\correspondingauthor{Jacob Pilawa \tiny{\citeColored{white}{Quennevilleetal2021a}}}
\email{jacobpilawa@berkeley.edu}

\author{Jacob D. Pilawa}
\affiliation{Department of Astronomy, University of California, Berkeley, CA 94720, USA}

\author{Christopher M. Liepold}
\affiliation{Department of Astronomy, University of California, Berkeley, CA 94720, USA}
\affiliation{Department of Physics, University of California, Berkeley, CA 94720, USA}

\author{Silvana C. Delgado Andrade}
\affiliation{George P. and Cynthia Woods Mitchell Institute for Fundamental Physics and Astronomy, and Department of Physics and Astronomy, \\
Texas A\&M University, College Station, TX 77843, USA}

\author{Jonelle L. Walsh}
\affiliation{George P. and Cynthia Woods Mitchell Institute for Fundamental Physics and Astronomy, and Department of Physics and Astronomy, \\
Texas A\&M University, College Station, TX 77843, USA}

\author{Chung-Pei Ma}
\affiliation{Department of Astronomy, University of California, Berkeley, CA 94720, USA}
\affiliation{Department of Physics, University of California, Berkeley, CA 94720, USA}

\author{Matthew E. Quenneville}
\affiliation{Department of Astronomy, University of California, Berkeley, CA 94720, USA}
\affiliation{Department of Physics, University of California, Berkeley, CA 94720, USA}

\author{Jenny E. Greene}
\affiliation{Department of Astrophysical Sciences, Princeton University, Princeton, NJ 08544, USA}

\author{John P. Blakeslee}
\affiliation{NSF's NOIRLab, Tucson, AZ, 85719, USA}

\begin{abstract}

We present a stellar dynamical mass measurement of a newly detected supermassive black hole (SMBH) at the center of the fast-rotating, massive elliptical galaxy NGC~2693 as part of the MASSIVE survey. We combine high signal-to-noise integral field spectroscopy (IFS) from the Gemini Multi-Object Spectrograph (GMOS) with wide-field data from the Mitchell Spectrograph at McDonald Observatory to extract and model stellar kinematics of NGC~2693 from the central $\sim 150$ pc out to $\sim2.5$ effective radii.   
Observations from \textit{Hubble Space Telescope} (\textit{HST}) WFC3 are used to determine the stellar light distribution.  We perform fully triaxial Schwarzschild orbit modeling using the latest TriOS code and a Bayesian search in 6-D galaxy model parameter space to determine NGC~2693's SMBH mass (\mbh), stellar mass-to-light ratio, dark matter content, and intrinsic shape.
We find $\mbh = \left(1.7\pm 0.4\right)\times 10^{9}\ \msun$ and a triaxial intrinsic shape with axis ratios $p=b/a=0.902 \pm 0.009$ and $q=c/a=0.721^{+0.011}_{-0.010}$, triaxiality parameter $T = 0.39 \pm 0.04$.
In comparison, \cpm{the best-fit  orbit model in the axisymmetric limit and (cylindrical) Jeans anisotropic model of NGC~2693 prefer $\mbh = \left(2.4\pm 0.6\right)\times 10^{9}\ \msun$ and $\mbh = \left(2.9\pm 0.3\right)\times 10^{9}\ \msun$, respectively. Neither model can account for the non-axisymmetric stellar velocity features present in the IFS data. }

\end{abstract}


\section{Introduction} 

The most massive SMBHs in the local universe have been found at the centers of some of the most massive nearby elliptical galaxies.  
\cpm{At stellar masses $M_* \gtrsim 10^{11.5} M_\odot$ targeted by the volume-limited MASSIVE galaxy survey \citep{Maetal2014}},
a majority of these massive elliptical galaxies
exhibit slow or no detectable rotation \citep{Vealeetal2017b, Vealeetal2017a, Eneetal2018}. \cpm{When selected by environments, massive galaxies in the cores of galaxy clusters are also predominantly slow- or non-rotators (e.g., \citealt{Broughetal2017, Loubseretal2018, Krajnovicetal2018b, Grahametal2018}).
In comparison, the ATLAS$^{\rm 3D}$ project surveyed early-type galaxies at lower masses ($10^{10} M_\odot \la M_\star \la 10^{11.5} M_\odot$) and found most to be fast rotators 
\citep{emsellemetal2011}.}

When the kinematic axis of a massive elliptical galaxy can be identified, it is often misaligned from the photometric major axis
(e.g., \citealt{Eneetal2018}). 
Detailed IFS kinematic maps also show intricate local twists, and the central and main-body kinematic axes within a galaxy are not always aligned \citep{Eneetal2020, Krajnovicetal2020}.
All these features are strong indications that local massive elliptical galaxies are triaxial in intrinsic shape, and not axisymmetric as is \cpm{often assumed in prior dynamical modeling studies of early-type galaxies}, which would only allow rotation about the minor axis \citep{Binney1985}. 

The MASSIVE galaxy survey \citep{Maetal2014} is 
designed to study all the major dynamical components
-- SMBH, stars, and dark matter halo -- in the most massive $\sim 100$
$\left(M_* \gtrsim 10^{11.5} M_\odot\right)$ 
early type galaxies (ETGs) in the local volume (out to $\sim 100$ Mpc). 
For a subset of 20 MASSIVE galaxies,
we have completed the stellar kinematic measurements from our IFS observations that cover both the galaxies' central regions with high spatial resolution, and wide fields out to at least one effective radius. In this paper, we focus on NGC~2693, a galaxy with one of the largest \cpm{ratios of $V/\sigma$, where the stellar rotation is $V \sim 160 \kms$ and the velocity dispersion is $\sigma \sim 300 \kms$ \citep{Vealeetal2017b, Eneetal2020}}.
\cpm{The only other MASSIVE galaxy that exhibits similarly fast and regular rotation is NGC~1453.}
NGC~1453 is recently studied both in the axisymmetric limit \citep{Liepoldetal2020} and with fully triaxial orbit modeling \citep{Quennevilleetal2021b}.
The triaxial models are found to better recover the input kinematics while also fitting the non-axisymmetric features present in NGC~1453. 
Given the similarities between NGC~2693 and NGC~1453, we turn to NGC~2693 in this paper.

We use the orbit superposition method to obtain
dynamical mass measurements of the components of NGC~2693 with observations taken as part of the MASSIVE Survey.
We use our latest version of the TriOS code \citep{Quennevilleetal2021a, Quennevilleetal2021b} \cpm{based on the code by \citet{vandenBoschetal2008}\footnote{https://github.com/remcovandenbosch/TriaxSchwarzschild}}
to perform a full triaxial modeling of the stellar orbits in NGC~2693 and to simultaneously constrain the galaxy's intrinsic shape, \mbh, and other mass parameters.

In Section~2, we describe the photometric observations used to model the deprojected stellar mass distribution of NGC~2693, as well as the spectroscopic observations from GMOS (\citealt{Hooketal2004}) covering the central kpc and wide-field observations from the McDonald Mitchell IFS (\citealt{hilletal2008a}). In Section~3, we describe the triaxial modeling and phase space sampling in our Schwarzschild orbit models. In Section~4, we discuss our search for the best-fit triaxial galaxy model, marginalization scheme for extracting best-fit parameters, and resulting best-fit dynamical model. In Section~5, we compare the triaxial model to axisymmetric Schwarzschild orbit models and Jeans modeling of NGC~2693. 

We adopt a distance to NGC~2693 of 71.0 Mpc from the MASSIVE-WFC3 project \citep{Goullaudetal2018} using the surface-brightness fluctuation technique \citep{Jensenetal2021, Blakesleeetal2021}. At this distance, $1''$ is 354 pc, assuming a flat $\Lambda$CDM cosmology with a matter density of $\Omega_m = 0.315$ and a Hubble parameter of $H_0 = 70\kms \text{ Mpc}^{-1}$.

\begin{figure*}
  \includegraphics[width=7in]{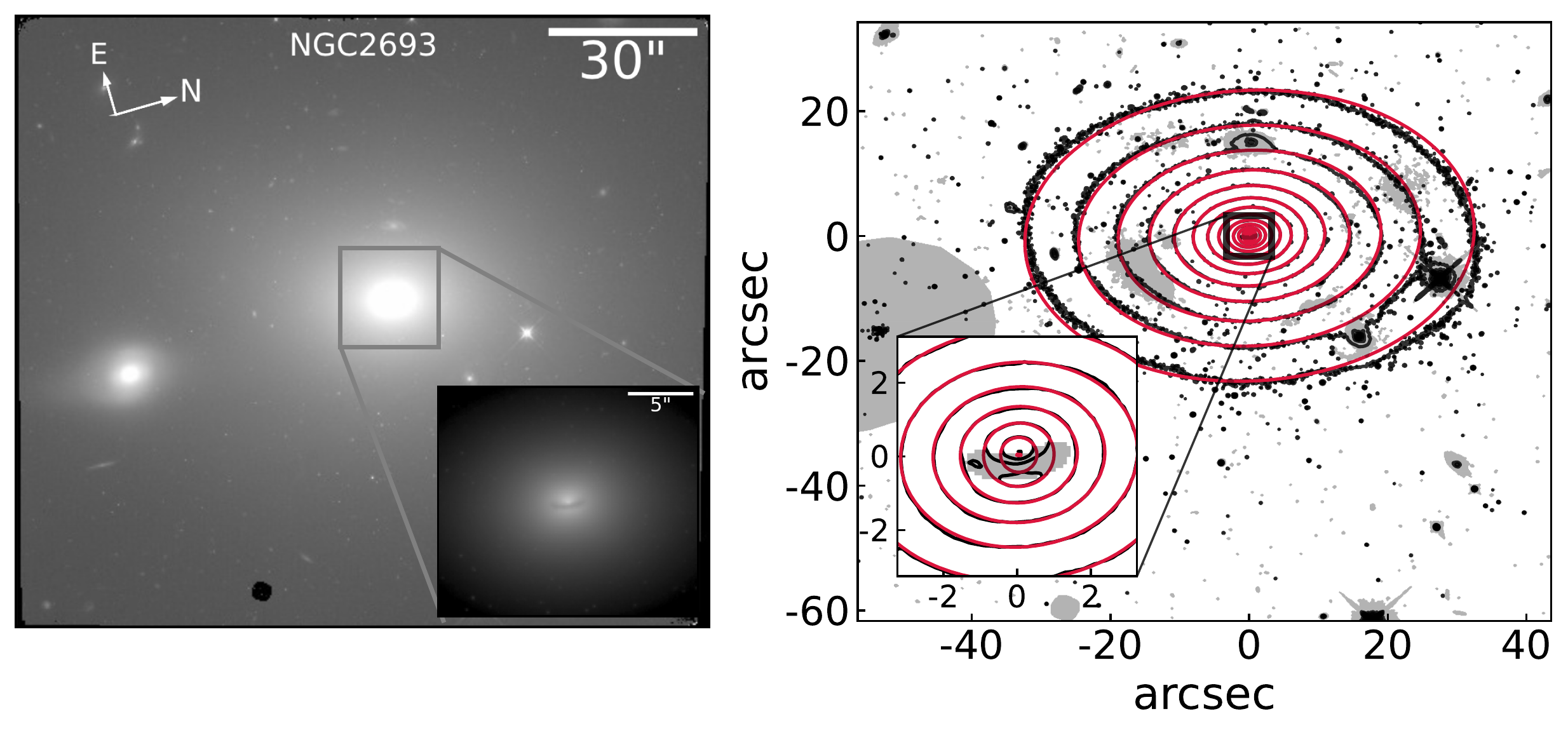}
    \caption{
    (Left) The F110W-band \textit{HST} image of NGC~2693 used for our photometry. A companion galaxy which is masked from photometric analysis (see text) can be seen $\sim 50''$ south of NGC~2693. (Left inset) NGC~2693 has a dust disk extending approximately $1.5''$ (in radius) from the center. This feature is masked from our MGE fitting. (Right) Isophotes of the \textit{HST} WFC3 IR image of NGC~2693 (black) and the best fit MGE model (red). The inset shows the central region containing the nuclear dust disk. The gray regions are masked when performing the fit as described in the text. 
    }
\label{fig:MGE}
\end{figure*}

\section{Photometric and Spectroscopic Observations}
\label{sec:two}

NGC~2693 is a relatively isolated galaxy, being the only identified member of its galaxy group in the 2MASS ``high-density contrast'' group catalog \citep{crooketal2007}.
We obtain photometric observations of NGC~2693 in the F110W filter of \textit{HST} and spectroscopic observations using GMOS in IFS mode on the 8.1 m Gemini North telescope and the Mitchell IFS on the 2.7m Harlan J. Smith Telescope at McDonald Observatory. In this section, we describe these observations, the data reduction process, modeling the surface brightness profile, and the extraction of the stellar kinematics of NGC~2693.  

\begin{table}[ht]
\centering
\caption{Best-fit MGE parameters to the \ngc\ \emph{HST} WFC3 IR photometry. Each Gaussian component ($k$) is parametrized by a central surface density $I_k = L_k/2\pi \sigma_k^{\prime 2} q_k^\prime$, dispersion $\sigma_k^{\prime}$, and axis ratio $q_k^{\prime}$, where primed variables are projected quantities. The central surface densities have been corrected for a galactic extinction of 0.017 mag and assume an absolute (Vega) magnitude of 3.89 for the Sun. The components all have a PA of 167.9$^\circ$ east of north.\\
}

\begin{tabular}{c|c|c}\hline
$I_k \ \ [L_{F110W,\odot} / {\rm pc}^{2}]$ & $ \ \ \sigma_k^{\prime} \ \ [''] \ \ $ & $ \ \ q_k^{\prime} \ \ $ \\
\hline
 $ 6521.78208  $   & $ \  0.10611 \ $       & $\ 0.99990 \ $ \\
 $18597.03132  $   &   $  0.24071   $       &   $0.99990   $ \\ 
 $17231.46878  $   &   $  0.51203   $       &   $0.98100   $ \\
 $14094.87758  $   &   $  1.21090   $       &   $0.78400   $ \\  
 $ 4883.63124  $   &   $  2.49551   $       &   $0.76540   $ \\
 $ 1854.85184  $   &   $  4.66292   $       &   $0.74880   $ \\
 $  694.17907  $   &   $  8.71737   $       &   $0.73490   $ \\  
 $  177.10426  $   &   $ 12.21259   $       &   $0.68370   $ \\
 $  217.66116  $   &   $ 21.51988   $       &   $0.71920   $ \\
 $   59.57494  $   &   $ 66.58575   $       &   $0.78070   $ \\
\end{tabular}
\label{tab:mge}
\end{table}

\subsection{HST Observations and Stellar Mass Profile of NGC~2693}

We model the spatial distribution of stars in NGC~2693 with observations from the IR channel of \textit{HST} WFC3 in the F110W filter (Figure~\ref{fig:MGE}).
Observations (GO-14219, P.I. J. Blakeslee) were taken over a single orbit and have a total exposure time of 2695 seconds. 
This orbit was divided into five dithered exposures with a sub-pixel dither pattern to improve measurements of the point-spread function (PSF). We reduce the images using STScI's standard reduction pipeline, a specialized Python program\footnote{https://github.com/gbrammer/wfc3} to correct for variable background levels, and the {\tt Astrodrizzle} package \citep{gonzaga12}. We additionally perform background subtraction, removing a neighboring galaxy located 55\arcsec\ to the south of NGC~2693, and construct a mask to exclude foreground stars, other galaxies, and detector artifacts. The final F110W image has a pixel scale of 0\farcs1. For details on the photometric data reduction, see \citet{Goullaudetal2018}.

The \textit{HST} observations of NGC~2693 show slightly boxy isophotes near the center of the galaxy, which become disky at radii larger than $\sim5$\arcsec. There is a small compact dust disk extending 1\farcs5 in radius from the center.
The galaxy's luminosity-weighted ellipticity is nearly constant with radius (beyond the region of the central dust disk) with a mean value $\langle \epsilon \rangle_L = 0.27\pm 0.002$ \citep{Goullaudetal2018}.
Below we parametrize the surface brightness of NGC~2693 as a sum of 2D Gaussians with a common center and PA.

We run the \citet{Cappellari2002} Multi-Gaussian Expansion (MGE) code with regularization to avoid flattened components that artificially restrict the range of inclination angles that can be used during the dynamical modeling. We then tweak the MGE solution using {\tt Galfit} \citep{peng02}. We set a lower boundary on each component's projected axis ratio, $q_k^\prime$, of $0.65$, which was determined from the previous regularized MGE run. The WFC3 PSF is accounted for by an empirical PSF constructed from extracting, summing, and re-normalizing 11 bright stars within the field of view.
We apply the mask of other objects in the field plus a mask for the central dust disk. Since no high-resolution multi-band imaging for NGC~2693 was available, we construct a dust mask by eye initially, conservatively flagging only the most obviously affected pixels, and then we adopt an iterative approach. After the {\tt Galfit} run converged, we examine the residual image and extend the dust mask to neighboring pixels with residuals above a selected threshold. We continually repeat the process, each time modifying the threshold, until achieving residuals at the $\sim5$\% level at the nucleus.

Our best-fit MGE is composed of 10 Gaussians with the central surface brightness $I_k$, projected dispersion $\sigma_k^\prime$, and $q_k^\prime$ given in Table~\ref{tab:mge}. The model is a good description of the data, as seen in Figure~\ref{fig:MGE}, with residuals below $\sim6$\% out to a radius of $\sim70$\arcsec.

All of the MGE components have the same PA of $167.9^\circ$ east of north, but we run {\tt Galfit} again allowing for the PA to vary between components. The initial guesses for the parameters are set to the best-fit MGE from Table~\ref{tab:mge}, and we use the same empirical PSF and mask. When allowing for PA twists, our best-fit MGE consists of three circular Gaussians at small radii.
The next five Gaussians have PA twists of $< 3^\circ$ relative to a PA of $167.9^\circ$ and the outer three components have smaller PAs of $\sim 155^\circ$. While we subtracted the companion galaxy and masked the remaining residuals, the outermost region of NGC~2693 likely remains contaminated by the companion galaxy causing the smaller PAs for the largest three MGE components.
Nevertheless, these two MGEs differed by less than 5\% at all radii for which we have 
kinematic data. Beyond 50\arcsec, the relative difference approaches $\sim10\%$ primarily due to the companion. 
The negligible impact of allowing for a PA twist does not seem to improve our fit but rather fits the remaining contamination of the companion galaxy, and thus we adopt the MGE model with a spatially constant PA.

\cpm{We further test the companion galaxy's impact on the measured photometric position angle by fitting subsections of the full \textit{HST} WFC3 image. We perform two additional fits, one to the central $20\arcsec\times 20\arcsec$ region and another to the central $40\arcsec\times 40\arcsec$ region as opposed to the full $141\arcsec\times 125\arcsec$ image. We use our fiducial MGE presented in Table~\ref{tab:mge} as initial guesses for the ten Gaussians, and we require the PA and center for all ten components to be the same. In both cases, the preferred photometric position angle changes negligibly compared to the fiducal case of $167.9^\circ$ E of N. When fitting only to the central $20\arcsec\times 20\arcsec$ region of the image, the preferred photometric PA is $168.6^{\circ}$ E of N, and when fitting to the $40\arcsec\times 40\arcsec$ region, the PA of the Gaussians is $168.2^\circ$ E of N.}

\subsection{Central GMOS kinematics} 
\label{central_kpc_kinematics}

We observed the central region of NGC~2693 in the 2016B semester with the two-slit mode of GMOS, providing a $5'' \times 7''$ ($\sim$1.7 kpc $\times\,2.4$ kpc) field-of-view composed of 1000 hexagonal lenslets, for a projected diameter of $0.2''$ per lenslet. In total, six science exposures were taken, each of 1200 seconds exposure time. The median seeing was $0.6''$ FWHM.  We used the R400-G5305 grating with the CaT filter for clean coverage from $7800-9330$ \AA. A $5'' \times 3.5''$ region of the sky, offset $1'$ from NGC~2693's central region was simultaneously observed.  The spectral resolution was determined from arc lamp lines for each lenslet, with a median value of 2.3 \AA\ FWHM.

\begin{figure}[htp]
  \includegraphics[width=\columnwidth]{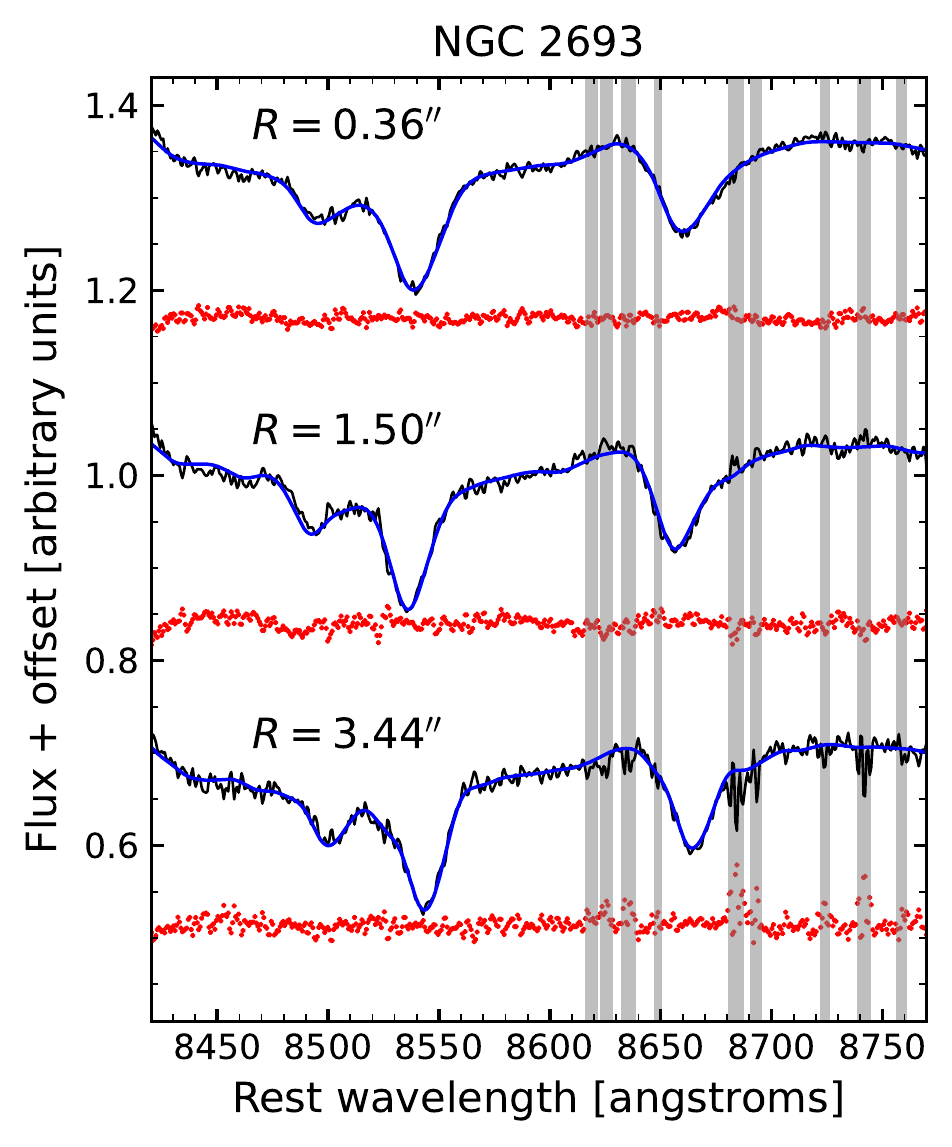}
    \caption{
    CaII triplet region of the GMOS IFS spectrum (black) of NGC~2693 for three example bins located at increasing distance from the nucleus: (top) central bin {with $S/N = 204$}, (middle) bin $1.50''$ from center {with $S/N = 144$}, and (bottom) bin $3.44''$ from center {with $S/N = 100$}. The spectrum is fit by broadening a series of stellar templates by the best-fit LOSVD (blue) over a wavelength range of $8420-8770$ \si{\angstrom}. The grey shaded regions are excluded from the fit to account for improperly subtracted sky lines. The red points are the (best-fit minus observation) residuals offset by constants for clarity. The typical residual is $\sim 0.5\%$. 
    }
    \label{fig:N2693_gmos}
\end{figure}

The GMOS lenslets are binned to achieve a target $S/N$ of 125 using the Voronoi-binning procedure \citep{CappellariCopin2003}. 
This procedure results in a total of 60 bins. 
The spectra are co-added from individual lenslets within a single Voronoi bin as described in \citet{Eneetal2019}. 
Example spectra for three bins at increasing radii are shown in Figure~\ref{fig:N2693_gmos} (black curves).

We use the CaII triplet absorption features over a rest wavelength of 8420-8770\AA\  
to extract the stellar line-of-sight velocity distribution (LOSVD) for each bin with the penalized pixel-fitting (pPXF) method of \citet{Cappellari2017}. 
We choose to decompose each LOSVD into a Gauss-Hermite series up to order $n=8$ as is done in \citet{Liepoldetal2020}. 
An additive polynomial of degree zero (a constant) and a multiplicative polynomial of degree three are used to model the stellar continuum for the spectra.

\begin{figure*}[htp]
  \includegraphics[width=6.5in]{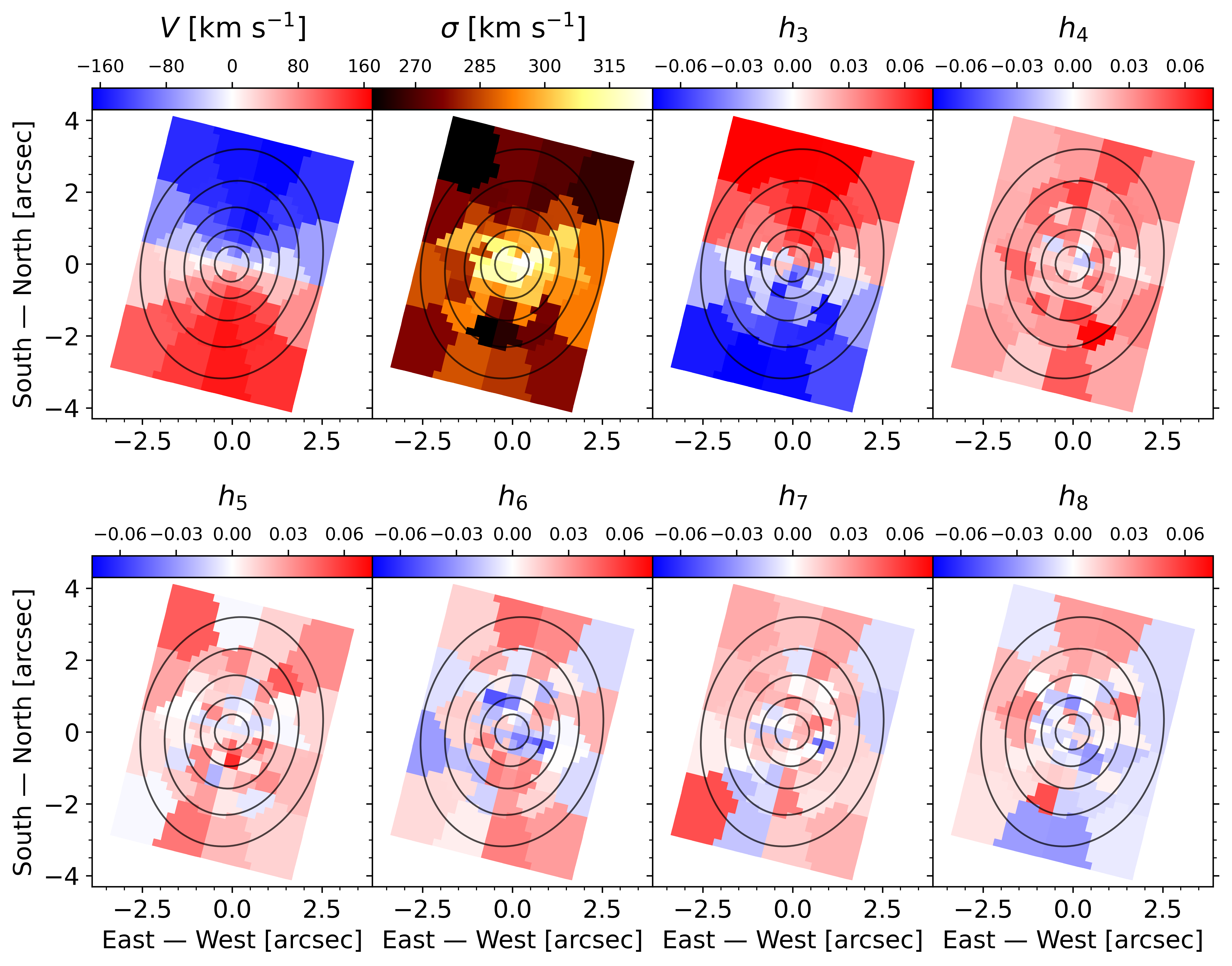}
    \caption{
    Spatial maps of the stellar kinematics extracted from the Gemini GMOS IFS spectra for 60 bins in the central $5''\times 7''$ region of \ngc. Each panel shows a different moment of the Gauss-Hermite expansion of the line-of-site velocity distribution: the top-left two panels are the velocity $V$ and velocity dispersion $\sigma$, with the higher-order $h_i$ moments characterizing deviations from Gaussianity. The $+x$ axis of the galaxy is located 167 degrees East of North (North is up and East is to the left). The velocity map shows a prominent rotation pattern with maximal velocities of $|V| \sim 160\kms$; the $\sigma$ maps shows a central peak.  \cpm{The black lines are surface brightness contours from the best-fitting MGE model in Table~\ref{tab:mge} and Figure~\ref{fig:MGE}}. 
    }
\label{fig:gmos_map}
\end{figure*}

\begin{figure*}[ht]
  \centering
  \includegraphics[width=\linewidth]{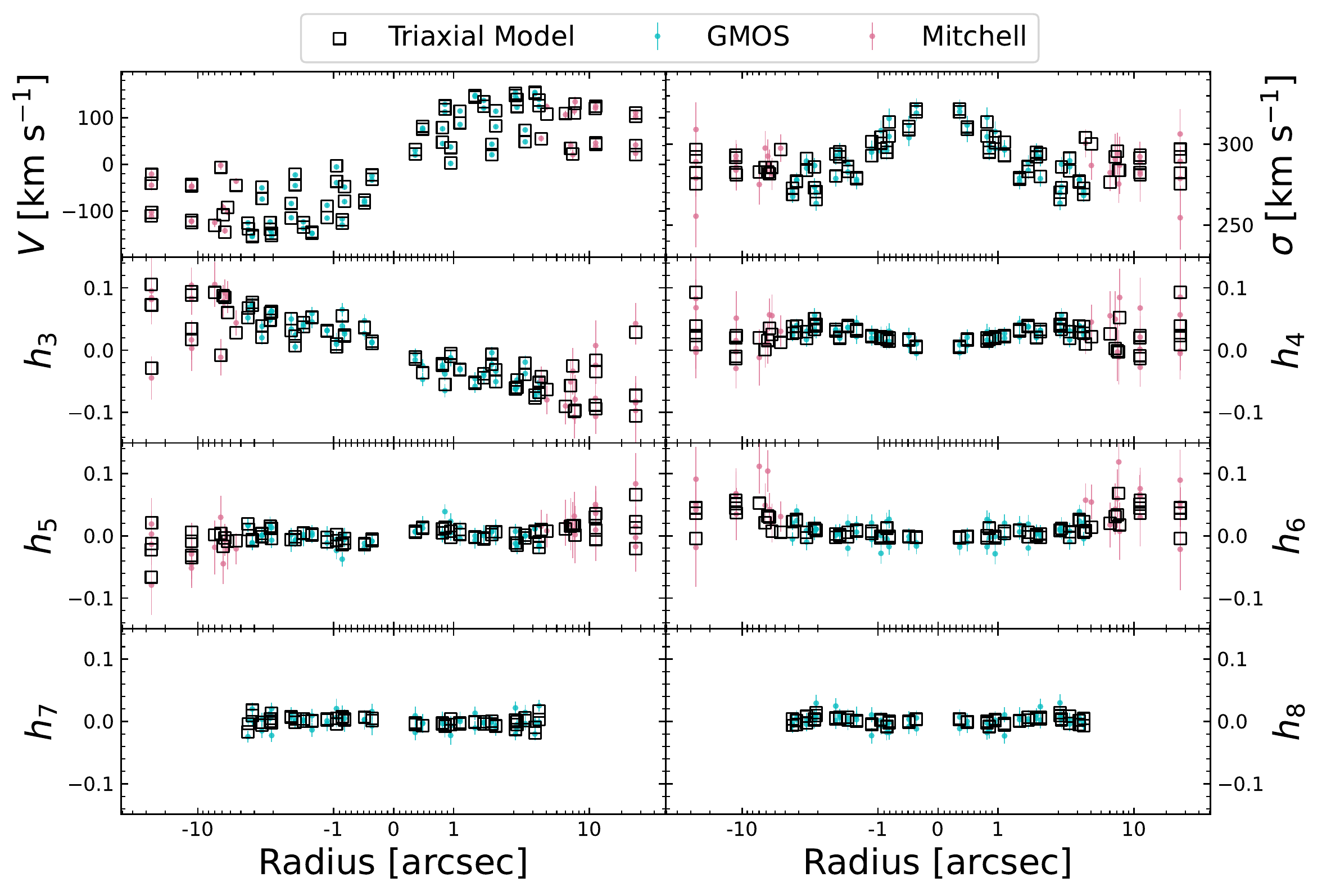}
  \hfill
  \caption{
  Radial profiles of the velocity moments determined from the GMOS (blue circles) and Mitchell (pink circles) IFS observations.  The observed moments are well matched by those predicted from our best-fit triaxial galaxy model (black open squares) with mass parameters $(\mbh, \ml, M_{15}) = (1.7\times 10^{9} \msun, 2.35, 7.1\times 10^{11} \msun)$ and shape parameters
  $(T, \tmaj, \tmin) = (0.39, 0.09, 0.17)$.  
  The spatial bins have been unfolded such that the bins whose centers lie between $-90^\circ$ and $+90^\circ$ of the photometric PA are plotted with positive radius and others with negative radius. Note that the axes are on a linear scale between $-1''$ and $1''$ and logarithmically scaled otherwise.}
\label{figure:triaxial_kinematics_radial}
\end{figure*}

We compare two sets of stellar templates chosen from the MILES Calcium Triplet (CaT) Library of 706 stars \citep{Cenarroetal2001}: the first set is limited to the 15 stars from Table~2 of \cite{Barthetal2002}; the second set includes all 706 stars in the library. The resulting kinematics are consistent within measurement errors, and we choose to use the former set of kinematics for
dynamical modeling discussed later in this paper.
These template spectra cover the wavelength range of 8347-9020\si{\angstrom} with a 1.5\si{\angstrom} spectral resolution FWHM.

We follow a nearly identical kinematic extraction procedure to that of \citet{Liepoldetal2020}.
Three example stellar templates broadened by the best-fit LOSVDs are shown in Figure~\ref{fig:N2693_gmos} (blue curves). The errors on the kinematic moments are determined with bootstrap methods as described in Section~4 of \cite{Eneetal2019}.
Figure~\ref{fig:gmos_map} shows the kinematic maps of the eight Gauss-Hermite velocity moments for all 60 GMOS bins;
Figure~\ref{figure:triaxial_kinematics_radial}  shows the corresponding radial profiles of the moments (blue filled circles).
The velocity map shows regular rotation with the maximum velocity reaching $|V| \sim 160\kms$; the $\sigma$ map shows a central peak with an amplitude of $\sim 320\kms$. The mean errors on $V$ and $\sigma$ are $4.3\kms$ and $5.0\kms$, respectively. The errors on higher order moments ($h_3$ through $h_8$) are similar in amplitude, ranging from $0.010$ to $0.013$.

\subsection{Wide-field Mitchell kinematics}

NGC~2693 is one of $\sim 100$ MASSIVE galaxies observed using the Mitchell IFS.
Three dither positions were used, and during each dither, we interleaved two 20 minute science frames with one 10 minute sky frame, for a total exposure time of 2 hours on-source. The Mitchell IFS consists of 246 fibers covering a $107'' \times 107''$ field of view. The observations covered a spectral range of 3650-5850\si{\angstrom} that include the Ca HK region, the $G-$band region, H$\beta$, Mg$b$, and several Fe features. 

Each fiber in the central region of NGC~2693 yields a spectrum of $S/N \gtrsim 50$, while the outer fibers are binned to achieve a $S/N\geq20$.  We obtain a total of 33 bins but drop the outermost 4 bins at $\sim 46''$ due to low $S/N$. We model the stellar LOSVD in a similar way to the GMOS data, fitting up to order $n=6$ due to the lower $S/N$. We use the MILES library of 985 stellar spectra (\citealt{Sanchez-Blazquezetal2006, Falcon-Barrosoetal2011}) as the templates. Details of the Mitchell data reduction and kinematic measurements are described in \citet{Vealeetal2017b} and \citet{Vealeetal2017a}.

The radial profiles of the Mitchell kinematic moments are shown in Figure~\ref{figure:triaxial_kinematics_radial} (pink filled circles).  The innermost few Mitchell fibers (each with a $4.1''$ diameter) overlap the FOV of GMOS. Reassuringly, the Mitchell kinematics in this region are in good agreement with the GMOS values averaged over the GMOS FOV.  The mean errors on the Mitchell kinematics are about two times larger than the GMOS kinematics. 
The mean errors on $V$ and $\sigma$ are $8.95\kms$ and $12.2\kms$, respectively, with errors on higher order moments ($h_3$ through $h_6$) ranging from $0.030$ to $0.039$.


\section{Triaxial Orbit Modeling of NGC 2693}
\label{sec:three}

\subsection{The TriOS Code and Input Kinematics}

We use the \textit{TriOS} code \citep{Quennevilleetal2021a,Quennevilleetal2021b}
to perform orbit modeling of NGC~2693. 
The galaxy is modeled as a stationary, triaxial gravitational potential composed of a SMBH, a stellar component specified by the deprojected MGE, and a dark matter halo.   The six parameter galaxy model and the method used to search this parameter space is described in detail in Section~\ref{subsec:galaxy_model}.

For a collection of orbits spanning the phase space, we integrate each orbit for 2000 (loop orbits) or 200 (box orbits) dynamical times. At a large number of steps along the trajectory, we project the orbit onto the sky, keeping track of the projected position and line-of-sight kinematics. For each dataset (GMOS and Mitchell), the projected position is convolved with the respective instrumental PSF, and steps in the trajectory are binned according to the apertures described in Sections 2.2 and 2.3. This produces (i) a measure of the fraction of that orbit's time spent in each aperture and (ii) the LOSVD associated with the orbits within each kinematic bin.  As described in Section~\ref{subsec:weights}, a superposition of these orbital contributions is found that best reproduces the observed kinematic LOSVD.
We repeat this process for many gravitational potentials to find the galaxy parametrization that most closely reproduces the observed kinematics.

We fit for eight moments of the GMOS kinematics and six moments of the Mitchell kinematics presented in Section~2.
We constrain additional moments up to and including $h_{12}$ to be $0.0 \pm \delta$, where $\delta$ represents typical errors in the highest odd and even moments of each kinematic dataset. 
\cpm{Our previous tests have showed that constraining only the lowest four moments and leaving the higher-order moments free in the orbit model could result in spurious behavior in the higher-order moments and the resulting LOSVDs (Figs.~10 and 11 of \citealt{Liepoldetal2020})}. 
We test the effects of constraining up to $h_{16}$, but moments $h_{13}$ to $h_{16}$ are sufficiently close to $0$ when constraining the first twelve moments that the added constraint on the last four moments does not change our fits. Thus, we opt to leave moments higher than $h_{13}$ unconstrained. We note that leaving moments $h_{9}$ to $h_{12}$ unconstrained can shift the inferred best fit parameters by $\sim 10\%$.

Throughout the analysis, we model the GMOS and Mitchell PSFs as single, circularly symmetric Gaussians with FWHM of $0.56''$ and $1.2''$ respectively. 
To keep the potential non-singular at the origin, we use a Plummer-style potential for the central black hole with a force softening length of $3\times 10^{-4}$ arcsec, which is roughly three orders of magnitude smaller than the central-most GMOS bins.

\subsection{Orbital Phase-space Sampling and Orbital Weight Optimization}
\label{subsec:weights}

Insufficient orbit sampling can bias the inferred mass and shape parameters, so particular care is needed in choosing initial conditions for the orbits.
In the TriOS code, 
the orbits are initialized in two different spaces, referred to as ``start spaces'' \citep{Schwarzschild1993,vandenBoschetal2008}. 
A Cartesian coordinate system centered on the galaxy's nucleus is used. The $x$, $y$, and $z$ axes are parallel to the intrinsic major axis $a$, intrinsic intermediate axis $b$,  and intrinsic minor axis $c$, respectively.

The first start space, called the $x$-$z$ start space, launches loop orbits from the $(x,z)$ plane and a velocity in the $+y$ direction at $N_E = 40$ different energies (implicitly sampled over radii).  The $(x,z)$ positions are determined by dividing the start space into $N_{I_2}$ rays spanning polar angles from $0$ to $\pi/2$ in the $x$--$z$ plane; along each ray, we space $N_{I_3}$ orbits. The code allows for additional dithering of orbits, where we group together $N_\text{dither} = 3$ adjacent initial conditions for denser and smoother phase-space sampling. Dithering orbits increases the sampling density by a factor of $N_\text{dither}^3$ since dithering is performed in all three dimensions. Orbits launched from the same initial position but with velocity in the $-y$ direction are valid orbits. To include these orbits in our modelling, we invert the LOSVD from each orbit in the $(x,z)$ start space and store the resulting orbits in a second `retrograde' library.

Similar to NGC~1453 (Section~4.3 of \citealt{Quennevilleetal2021b}),
we find spurious oscillations in the goodness-of-fit $\chi^2$ landscape while varying $T$ for NGC~2693 when using $N_{I_2} = 9$ and $N_{I_3} = 9$ in the $(x,z)$ start space. The spacing between these oscillations matches the spacing between dithered initial conditions, resulting in periodic local minima and thus biased results for the intrinsic galaxy shape. 
We eliminate these unwanted oscillations by increasing the angular sampling $N_{I_2}$ of orbits in the $(x,z)$ start space from $N_{I_2} = 9 $ to $N_{I_2} = 15$. 
The total number of orbits used in our loop library is therefore $2\times N_\text{dither}^3 \times N_E \times N_{I_2} \times N_{I_{3}} = 2\times3^3\times40\times15\times9 = 291,600$, where the factor of $2$ accounts for the time reversed copy of each orbit.

A second start space, called the stationary start space, is used to generate a library of box orbits in a triaxial system.  In this  start space, orbits with a given energy $E$ are launched at rest with $v_x = v_y = v_z = 0$ from starting positions, where the gravitational potential is $\Phi(R,\theta,\phi) = E$, and $\theta$ and $\phi$ are the polar and azimuthal angles in the usual spherical coordinate system. Both $\theta$ and $\phi$ are sampled uniformly between $0$ and $\pi / 2$ at $N_{\theta} = 9$ and $N_\phi = 9$ locations. We find no oscillatory behavior in the resulting $\chi^2$ for this start space, so we choose $(N_{\theta}, N_{\phi}, N_{dither}) = (9,9,3)$. The total number of orbits in our box library is thus $3^3 \times 40 \times 9 \times 9 = 87,480$. 

With three integrated orbit libraries consisting of $291,600+87,480=379,080$ orbits for a given galaxy model, we solve for the linear combination of orbital weights that best fits the observed kinematics and surface brightness. As noted, we use a dithering factor of 3, meaning that $3^3 = 27$ neighboring orbits are bundled while solving for the orbital weights. Accordingly, there are $379,080 / 3^3 = 14,040$ independent weights in each model. We compute these weights to minimize the $\chi^2$ associated with the kinematics using non-negative least squares \citep{LawsonHanson1995}, under the constraint that both the projected mass within each aperture and the 3D mass distribution in coarse bins are consistent within 1\% of the MGE. 
We do this for all kinematic bins simultaneously to build a model LOSVD in each bin for each galaxy model. 

\cpm{We note that the choice of 1\% mass constraint above is based on the range of values commonly adopted in earlier orbit modeling work (e.g., 2\% in \citealt{vandenboschdezeeuw2010}; 1\% in \citealt{Walshetal2017}). Our results do not depend on the exact value used: relaxing the 3D mass constraint from 1\% to 10\% changes our best-fit \mbh\ reported below by only a few \%, which is well within the $1\sigma$ uncertainty. Additionally, the typical $\chi^2$ associated with the mass constraint is only a small fraction ($\sim0.5\%$) of the size of the $\chi^2$ from the stellar kinematics.  We obtain identical \mbh\ regardless of whether we use $\chi^2$ from kinematics alone or include $\chi^2$ associated with the mass constraint.}

\section{Parameter Search of Triaxial Models}
\label{sec:four}

\subsection{Galaxy Model}
\label{subsec:galaxy_model}

We use a six-parameter model to describe the triaxial potential of NGC~2693.
Three parameters are for the mass components: black hole mass $\mbh$, stellar mass-to-light ratio $M^*/L_{F110W}$ (hereafter $\ml)$, and 
enclosed dark matter mass at $15$ kpc, $M_{15}$.  A logarithmic dark matter halo is assumed, where the enclosed dark matter mass at radius $r$ is
\begin{equation}
\label{eq:loghalo}
    M(<r)=\frac{V_{c}^{2}}{G} \frac{r^{3}}{r^{2}+R_{c}^{2}} \,,
\end{equation}
where $R_c$ is the scale radius and \vc is the circular velocity of the halo.
We fit for the enclosed dark matter mass at 15 kpc, which is approximately the outer edge of our outermost mass bin.

The other three parameters of our galaxy model encode the intrinsic shape of NGC~2693. 
Previous dynamical studies used angles $(\theta, \phi, \psi)$, or axis ratios $(u, p, q) = (a^\prime/a, b/a, c/a)$, to relate the projected and intrinsic shapes of a triaxial galaxy.  Here, $u$ is a compression factor relating the intrinsic (unprimed) and projected (primed) length scales;
$p$ is a ratio between the intrinsic intermediate axis $b$ and the intrinsic major axis $a$ (with projected major axis $a^\prime$); and $q$ is the ratio between the intrinsic minor axis $c$ and the intrinsic major axis.
Additionally, $q^\prime = b^\prime / a^\prime$ 
describes the projected flattening of the MGE component. In general these axis ratios are different for each MGE component as those components have different projected flattenings.

We instead adopt the new parameters $(T, \tmaj, \tmin)$ introduced in \citet{Quennevilleetal2021b}:
\begin{equation}\label{eqn:ShapeParams}
  T =\frac{1-p^{2}}{1-q^{2}}\,, \quad T_{\mathrm{maj}} =\frac{1-u^{2}}{1-p^{2}}\,, \quad
  T_{\min } =\frac{\left(u q^{\prime}\right)^{2}-q^{2}}{p^{2}-q^{2}} \,,
\end{equation}
where $T$ is the commonly used triaxiality parameter, $\tmaj$ parametrizes the length of the projected major axis 
relative to the minimum value $b$ and maximum value $a$, and $\tmin$ parametrizes the length of the projected minor axis
relative to the minimum $c$ and maximum $b$. 
The three parameters $T, \tmaj$ and $\tmin$ form a convenient unit cube, each with an allowed range of 0 to 1.  They have a number of additional desirable properties when compared to the axis ratios $(u, p, q)$ or angles $(\theta, \phi, \psi)$; see Section~3.4 of \citet{Quennevilleetal2021b}. 
Note that while these shape parameters are expressed in terms of the axis ratios in Equation~(\ref{eqn:ShapeParams}), they are constant for different MGE components when the PAs of the MGE components are identical. As shown in Equations~(8) and (A2) in \citet{Quennevilleetal2021b}, the angles $(\theta, \phi, \psi)$ can be directly computed from a $(T, \tmaj, \tmin)$ triplet and vice versa.

\subsection{Latin Hypercube Sampling and Bayesian Search}
\label{subsec:LHS}

As in \citet{Quennevilleetal2021b}, we
use the grid-free Latin hypercube sampling method \citep{McKay1979} to search the 6D model parameter space: \mbh, \ml, $M_{15}$, $T$, \tmaj, and \tmin. 
Latin hypercube sampling is becoming increasingly common in computer-designed experiments due to the simplicity of the sampling algorithm and the desirable space-filling properties in high dimensional parameter spaces. We adopt the Latin hypercube scheme described in \citet{Jin2005} and implemented in the \texttt{Python} package \texttt{SMT} \citep{SMT2019}.

In this method, we first divide each dimension of our search-space into $N$ cells, where $N$ is the number of galaxy models in a Latin hypercube batch. 
We then uniformly sample points in each dimension until each cell contains a point, uniformly filling the space. We note that we use the ``center'' dispersal criterion in \texttt{SMT}, which places new points in the center of each hypercube cell.
The result is a set of model points spanning six dimensions more uniformly than a regular grid and allowing for a more representative sampling of the likelihood function as a function of the six model parameters.

\begin{figure*}[ht]
  \centering
    \includegraphics[width=6in]{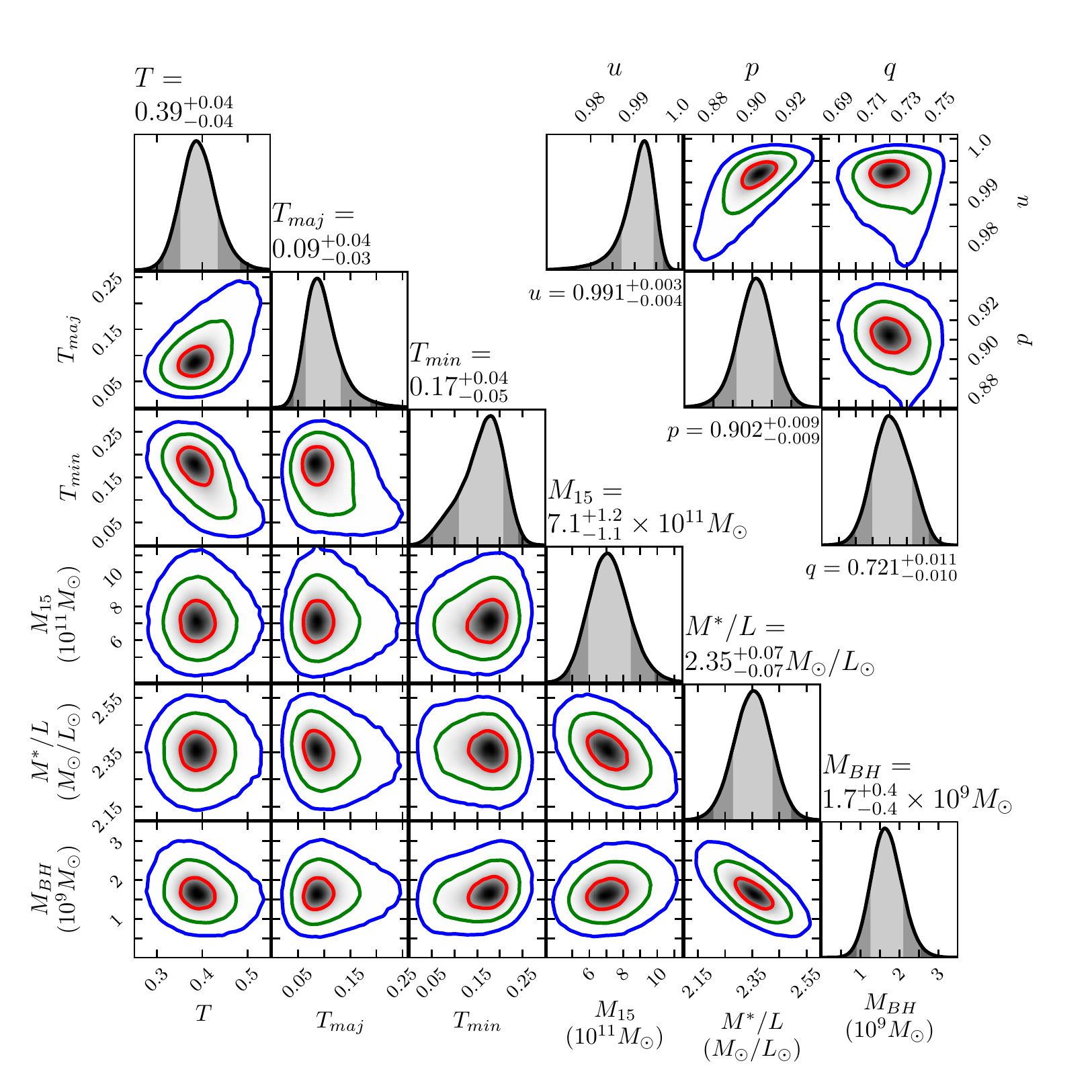}
  \caption{(Lower left) 1D and 2D marginalized posteriors for the triaxial orbit models of NGC~2693 described in the text. We marginalize over a smoothed 6D landscape generated with Gaussian process regression. The $1\sigma$, $2\sigma$, and $3\sigma$ contours are represented by the curves in red, green, and blue respectively, and as different shade of grey in the 1D panels. Above each 1D posterior distribution are the extracted best-fit values and $1\sigma$ confidence intervals. (Upper right) 1D and 2D marginalized posteriors in the axis-ratio space of $(u,p,q)$. Below each 1D posterior are the best-fit values and 1$\sigma$ confidence intervals. 
  } \label{fig:CornerPlots}
\end{figure*}

When drawing points from intrinsic-shape space, we opt to sample uniformly in $\sqrt{\tmaj}$ and $\sqrt{\tmin}$ rather than in $\tmaj$ and $\tmin$. For nearly axisymmetric galaxies, this sampling space results in fewer unrealistically flat models, increasing the efficiency of our parameter search.

After running preliminary searches over broad ranges of parameters, we choose the uniform prior ranges of $\mbh \in [0,4]\times 10^{9} \msun$, $\ml \in [2.0, 2.85]\ M_\odot / L_\odot$, $M_{15} \in [1,13]\times 10^{11} \msun$, $T \in [0.15,0.65]$, $\tmaj \in [0.0, 0.6]$, and $\tmin \in [0.0,0.4]$ for our parameters. 
After each hypercube realization of roughly $1000$ galaxy models, we model the $\chi^2$ likelihood landscape as a function of the six parameters using Gaussian process regression. We construct posterior distributions of the space and estimate of the best-fit values for each parameter using dynamic nested sampling \citep{Speagle_2020}. To test for convergence in our model sampling, we perform jackknife resampling where the regression and parameter inference is repeatedly performed using subsets of the full suite of models. In total, we generate 8530 galaxy model points which, when jackknife-resampled, converge on the same best-fit galaxy parameters.

To map the high-likelihood region in fine detail and to serve as an additional test for convergence from our hypercube iterations, we perform one more independent check of our best-fit parameters. 
We again perform a 6-D gaussian process regression function to the $\chi^2$ landscape produced by the 8530 models described above. We then sample 1000 additional model points with another hypercube, rejecting those which fall outside of the $3\sigma$ confidence volume, as estimated by that regression 
fitted to the previous 8530 model points. 
The rejection sampling procedure very efficiently populates the $\chi^2$ minimum. Of the 1000 points in that sample, $\sim9\%$ lie within the $3\sigma$ confidence volume for 6 parameters ($\Delta \chi^2 \approx 20.06$), compared to just $\sim4\%$ from the uninformed 8530-point sample. 

Both the uniform hypercube of 8530 models and the 1000 rejection-sampled hypercube models converge on the same best fit parameters. We include both sets here.
The resulting 6D posterior distribution
of our 9530 production run models is shown in Figure~\ref{fig:CornerPlots}. 
We determine the best-fit value and uncertainties by fitting the $\chi^2$ landscape with Gaussian process regression with a squared-exponential covariance kernel and sampling that landscape with the dynamic nested sampler described in \cite{Speagle_2020}. A uniform prior is assumed for all parameters.
The $1\sigma, 2\sigma,$ and $3\sigma$ confidence regions in 1D and 2D are computed from the posterior distribution marginalized over all other dimensions. These confidence levels correspond to $\Delta \chi^2 = 1, 4,$ and $9$ in 1-D and $\Delta \chi^2 \approx 2.3, 6.2, 11.8$ in 2D.

\subsection{Best-Fit Triaxial Model}

The best-fit galaxy model is an excellent fit to the observed stellar kinematics, shown in Figure~\ref{figure:triaxial_kinematics_radial}.
The best-fit parameters from the 6D posterior distribution in Figure~\ref{fig:CornerPlots} are listed in Table~\ref{tab:triax_results}.
The total $\chi^2$ of the best-fit triaxial model is $265.9$, with $19.7$ from the higher-order moments which are constrained to be $0.0 \pm \delta$. 
\cpm{As a test, we have repeated the regression using only four Gauss-Hermite moments measured from the spectra while setting $h_5$ and beyond to $0.0\pm \delta$ described above.  The best-fit \mbh\ is within $0.5\sigma$ confidence interval of the fiducial model in Table~\ref{tab:triax_results}, but the  uncertainties on \mbh\ in this case increase by $\sim10\%$. The same trend is reported in Table~2 of \citet{Liepoldetal2020}.}

\begin{table}[ht]
\noindent
\hspace{-5.5em}
\centering
\scalebox{0.9}{
\begin{tabular}{|l|l|l|l|}
\hline
\vtop{\hbox{\strut Galaxy}\hbox{\strut Parameter}}                    & \vtop{\hbox{\strut Triaxial}\hbox{\strut Orbit Model}}                  & \vtop{\hbox{\strut Axisymmetric}\hbox{\strut Orbit Model}}           & \vtop{\hbox{\strut JAM}\hbox{\strut Model}}           \\ \hline
$\mbh$ [$10^{9} M_\odot$]    & $1.7\pm0.4$               & $2.4\pm0.6$            & $2.9\pm0.3$   \\
$\ml$ [$M_\odot / L_\odot$]  & $2.35 \pm 0.07$           & $2.27 \pm 0.1$         & $2.17\pm0.03$ \\
$M_{15}\ [10^{11} M_\odot]$  & $7.1^{+1.2}_{-1.1}$       & $7.9 \pm 1.3$          & $4.7\pm0.2$   \\
$\beta_z$                    & See caption.$^\dagger$    & See caption.$^\dagger$ & $0.07\pm0.01$ \\
$T$                          & $0.39 \pm 0.04$           &                        &               \\
$\tmaj$                      & $0.09^{+0.04}_{-0.03} $   &                        &               \\
$\tmin$                      & $0.17^{+0.04}_{-0.05}$    &                        &               \\
$u$                          & $0.991^{+0.003}_{-0.004}$ &                        &               \\
$p$                          & $0.902 \pm 0.009$         &                        &               \\
$q$                          & $0.721^{+0.011}_{-0.010}$ &                        &               \\
$\theta \left(^\circ\right)$ & $66^{+4}_{-3}$            &                        &               \\
$\phi \left(^\circ\right)$   & $72\pm3$                  &                        &               \\
$\psi \left(^\circ\right)$   & $93.0^{+0.7}_{-0.6} $     &                        &               \\ \hline
\end{tabular}
}
\caption{Summary of best-fit galaxy models for NGC~2693. For each parameter, we marginalize over the other dimensions and report the 1$\sigma$ uncertainties. The axisymmetric orbit models and JAM models have fixed inclination of $70^\circ$. \cpm{In orbit models, $\theta$ is the inclination angle in the oblate axisymmetric limit ($\psi=90^\circ$, or equivalenly $p=1$), 
with $\theta=90^\circ$ being edge-on and $\theta=0^\circ$ being face-on}.
$^\dagger$We measure $\beta_z$ in the orbit model as a function of radius, shown in the bottom panel of Figure~\ref{fig:orbital_comp}. The best-fit JAM value of $\beta_z = 0.07\pm 0.01$ is consistent with the range of $\beta_z$ values measured from this best-fit model, with values ranging from $\beta_z = -0.27$ at small radii to $\beta_z = 0.28$ at large radii in both the triaxial and axisymmetric Schwarzschild models.}
\label{tab:triax_results}
\end{table}

As we will discuss further in Section~6.1, our inferred $\mbh = (1.7\pm 0.4) \times 10^9 M_\odot$ for the SMBH in NGC~2693 is within the intrinsic scatter of the SMBH-galaxy scaling relations.  For the intrinsic shape of NGC~2693, we can compare our inferred axis ratios $(p, q) = (0.902\pm 0.009, 0.721^{+0.011}_{-0.010})$ 
with the mean galaxy shapes for 49 slow-rotating galaxies in the MASSIVE survey from
\citet{Eneetal2018}.  In that work,
based on the observed ellipticity and misalignment between the kinematic and photometric axes, the mean flattening of the galaxy sample was estimated statistically to be $(p, q) = (0.88, 0.65)$, with $56\%$ of galaxies having $p>0.9$. 
The triaxial shape of NGC~2693 is therefore quite close to the mean values. NGC~1453, the other MASSIVE galaxy for which we have performed triaxial orbit modeling \citep{Quennevilleetal2021b}, has best-fit shape parameters of $(p, q) = (0.93, 0.78)$, indicating a slightly less flattened shape than NGC~2693 and the mean MASSIVE galaxy.  Additionally, the intrinsic shapes of NGC~2693 and NGC~1453 are consistent with the distribution of shapes of fast rotators found in the IllustrisTNG50 and IllustrisTNG100 simulations \citep{Pulsoni2020}, where the mean axis ratios are $(p,q)\sim(0.9,0.52)$ and the dispersion is $\sigma \sim 0.15$ for the most massive fast rotating elliptical galaxies.

\cpm{We note that by construction, the axis ratios $(p, q, u)$ of each MGE component obey the relation $0 \le q \le u q' \le p \le u \le 1$ (see, e.g., Sec.~2.1 of \citealt{Quennevilleetal2021b}).  For the MGE of NGC~2693, the most flattened component has $q' = 0.684$. For physically useful deprojections, we may expect $q \gtrsim 0.2$ \citep{binneydevacouleurs1981}. The allowed ranges of $(p, q, u)$ are therefore quite narrow, in particular for $u$, which is constrained to be between $\sim 0.9$ and 1. The errors on these parameters in Table~2, while appearing small on absolute terms, are on the order of $\sim 5-10$\% of the allowed ranges.}

\cpm{We also highlight that the recovered viewing angle $\theta = {66^\circ }^{+4^\circ}_{-3^\circ}$, which corresponds to the galaxy's inclination in the oblate axisymmetric limit, is consistent with the inclination of the galaxy estimated from the nuclear dust disk at NGC~2693's center, which we measure to be $i \approx 70^\circ$. }

\begin{figure}[htp]
  \includegraphics[width=\columnwidth]{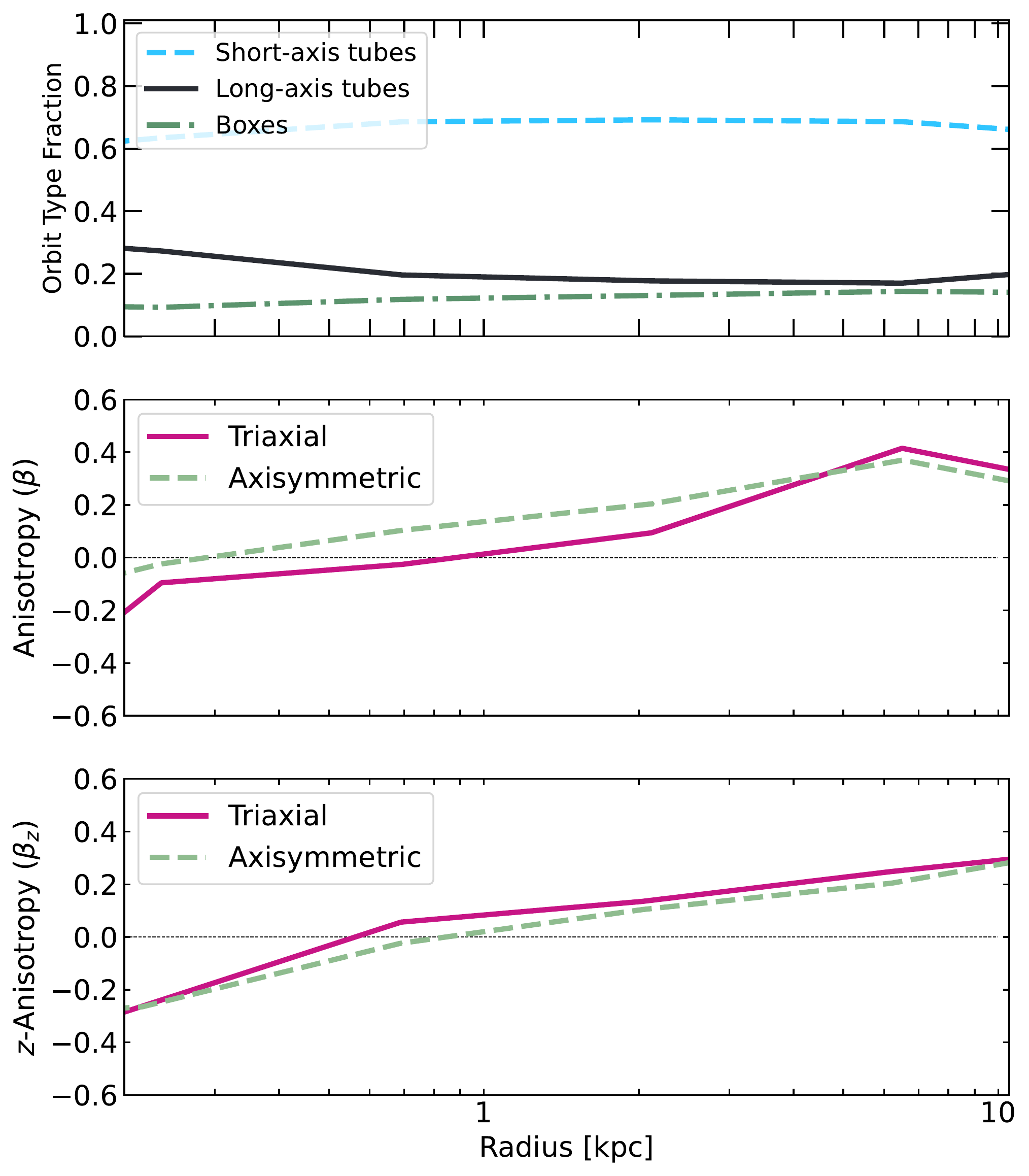}
    \caption{ 
(Top) Fraction of orbital weights in each orbital family for the best-fit triaxial galaxy model. The orbital structure is dominated by short-axis tubes at all radii, with a non-zero fraction of the weights occupied by long-axis tubes and box orbits, both of which are present only in triaxial potentials. For axisymmetric models, short-axis tubes are the only allowed orbital family. (Middle) Velocity anisotropy $\beta \equiv 1 - \sigma_t^2 / \sigma_r^2$ profile of the best-fit triaxial model (pink, solid line) and best-fit axisymmetric model (green, dashed line) of NGC~2693. Inner orbits are tangential out to $\sim 1 \text{ kpc}$ and are increasingly radially anisotropic at larger radii in both the axisymmetric and triaxial cases. (Bottom) Anisotropy parameter $\beta_z = 1 - (\sigma_z/\sigma_R)^2$, where $\sigma_z$ and $\sigma_R$ are the velocity dispersions parallel to the rotation axis and in the radial direction, for the best-fit axisymmetric and triaxial models described in the text. 
    }
    \label{fig:orbital_comp}
\end{figure}

\cpm{We have also run two additional galaxy models using the best-fit parameters shown in Table~\ref{tab:triax_results}, but with $\mbh = 0$ and $\mbh$ twice the best-fit value to assess what features in the kinematics provide the black hole mass constraint. Both models are a worse fit to the kinematic data, giving a $\Delta \chi^2 = 38.1$ when there is no black hole present, and a $\Delta \chi^2$ of $17.9$ when the black hole is twice as massive. 
As expected, the inner kinematic data provide significant constraints on \mbh, with $\sim 30\%$ and 50\% of the additional $\Delta \chi^2$ coming from the inner $\sim 1''$ data for the two test cases, respectively.}

We use the computed orbit libraries to calculate the orbital composition, as well as the radial velocity dispersion $\sigma_r$ and tangential velocity dispersion $\sigma_t\equiv\left(\sigma_\theta^2 + \sigma_\phi^2\right)/2$ of NGC~2693. We present the orbital fractions and two anisotropy parameters $\beta$ and $\beta_z$ as a function of radius in Figure~\ref{fig:orbital_comp}. 
Long-axis tubes and box orbits, both of which are only present in triaxial potentials, make up $\sim40\%$ of the orbits at small radii and $35\%$ of the orbits at outer parts of the galaxy. 
Near the center of the galaxy, the orbits are mostly tangential with $\beta < 0$ but become radially anisotropic beyond $\sim1\text{ kpc}$.

\section{Axisymmetric Dynamical Modeling of NGC~2693}
\label{sec:five}

For a comparison study, we have performed axisymmetric modeling of NGC~2693 using both the orbit superposition method and Jeans modeling.  We describe the results from each method below.

\subsection{Schwarzschild Orbit Modeling in the Axiysmmetric Limit}
\label{subsec:axi_orbit_modeling}

We use the axi-symmeterized version of the TriOS code first described in \citet{Quennevilleetal2021a}, with further improvements in mass binning and acceleration table discussed in \citet{Quennevilleetal2021b}.
\citet{Liepoldetal2020} first applied this code to NGC~1453; here we use similar settings \cpm{to achieve axisymmetry within the triaxial TriOS code}.
We ensure the low $L_z$ space is well sampled by tube orbits and do not include the box orbit library (which has $L_z=0$) explicitly. We set the viewing angle $\psi$ sufficiently close to $90^\circ$ in the input parameter file, i.e., $|\psi - 90^\circ| = 10^{-9}$, to ensure no long-axis tube orbits are present.
For the remaining short-axis tube orbits,
we enforce axisymmetry by making 40 copies of each orbit, each copy rotated successively by $2\pi/40$ about the intrinsic minor axis of the galaxy (Section~3 of \citealt{Quennevilleetal2021a}).   These three precautions are necessary to run the triaxial code in the axisymmetric limit and obtain robust parameter constraints.
We choose $(N_{I_2}, N_{I_3}, N_{dither}) = (9,9,3)$ for the phase space sampling, and include two copies of the integrated orbit library in our minimization. 
This gives a total number of $174,960$ orbits for our axisymmetric galaxy models. 

We search for the best-fit galaxy model using the Latin hypercube scheme outlined in Section 3 over three dimensions: \mbh, \ml, and $M_{15}$, with a fixed inclination angle $i = 70^\circ$, estimated from the nuclear dust disk. Our hypercube consists of 2000 galaxy models drawn from a range $\mbh = [0,4]\times 10^{9}\ \msun$, $\ml = [1.8, 2.8]\ M_\odot / L_\odot$, and $M_{15} = [1,13]\times 10^{11}\ \msun$. 
The best-fit axisymmetric model parameters are listed in Table~\ref{tab:triax_results}.

\begin{figure}[htp]
  \includegraphics[width=\columnwidth]{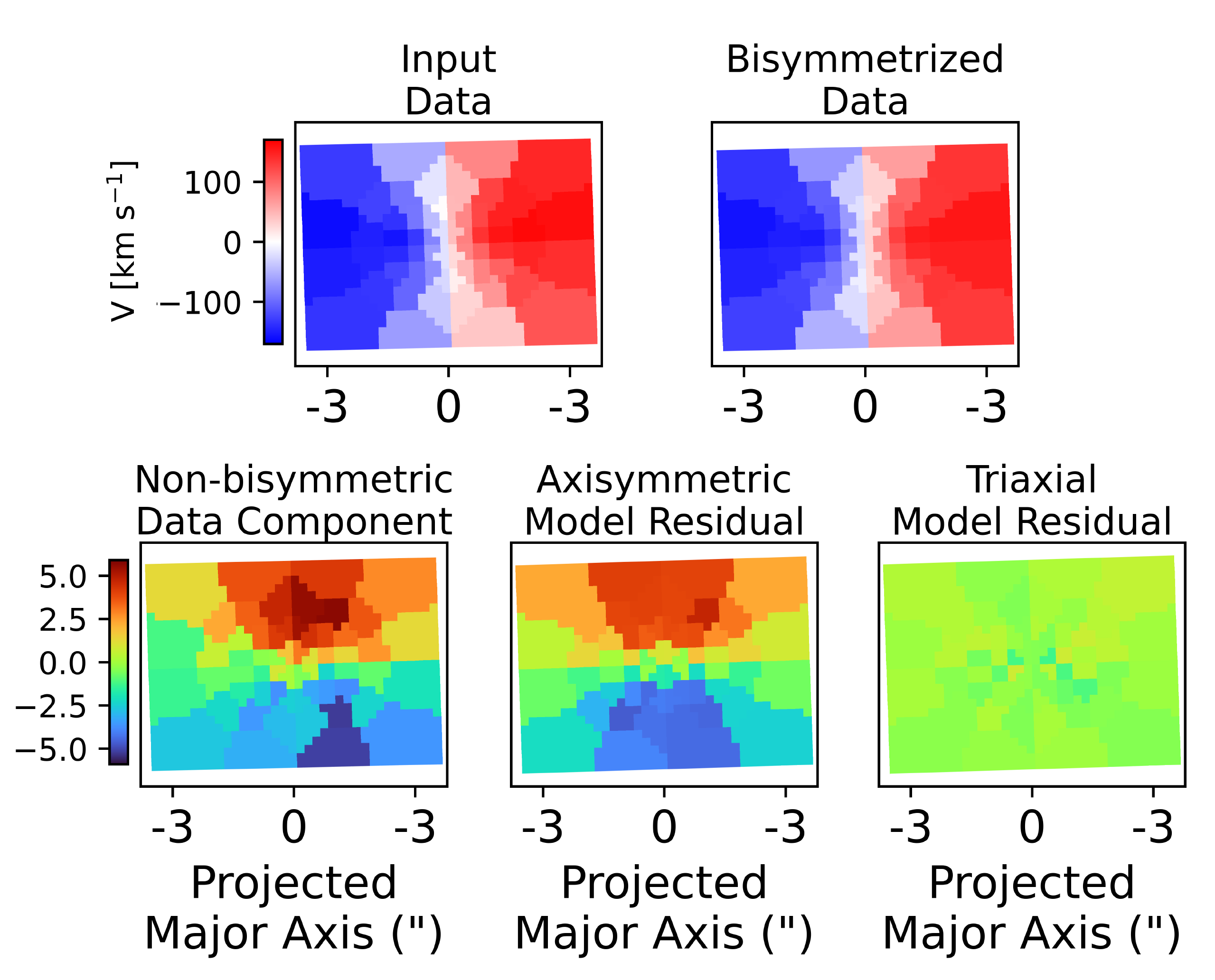}
    \caption{
    Observed velocity map of the central $5''\times 7''$ of NGC~2693 (upper left), oriented such that the observed photometric major and minor axes are horizontal and vertical, respectively, where the best-fit MGE PA is $167.9^\circ$.  We decompose the map 
    into a bisymmetrized component (upper right) and a non-bisymmetric component (lower row),
    where bisymmetry means  
    symmetry for points mirrored about the photometric major axis and anti-symmetry for points mirrored across the photometric minor axis.  
    The non-bisymmetric component (normalized by measurement uncertainty) shows a prominent apparent minor-axis rotation, a telltale sign of triaxiality.  
    Since axisymmetric models can only produce bisymmetric velocity maps by construction, the residuals from our best-fit axisymmetric model (lower middle) shows a similar pattern as the non-bisymmetrized map. By contrast,
    our best-fit triaxial model (lower right) is able to reproduce the full observed velocity structure, and the residuals scatter randomly about $0$.
    }
    \label{fig:bisymmetric_decomposition}
\end{figure}

Our best fit axisymmetric model of NGC~2693 prefers a $\sim40\%$ larger \mbh\ compared to the triaxial case, though the recovered best-fit \ml\ and dark matter halo are consistent with triaxial modeling at the $1\sigma$ level.
By construction, axisymmetric models produce bisymmetric kinematic maps, meaning that the LOSVDs are symmetric about the photometric major axis and anti-symmetric for points mirrored about the photometric minor axis. 
By contrast, LOSVDs in triaxial models are only point-symmetric about the origin. 
The apparent minor-axis rotation in ``Non-bisymmetric Data Component" panel of Figure~\ref{fig:bisymmetric_decomposition} therefore can be fit by triaxial models but not axisymmetric models.
Our best-fit axisymmetric model fails to account for this component (lower middle panel), while
our best-fit triaxial model captures well the full velocity features and produces featureless and nearly zero residuals (lower right panel).
Our best-fit axisymmetric model fails to account for this component (middle, left panel), while
our best-fit triaxial model captures well the full velocity features and produces featureless and nearly zero residuals (middle, right panel).

\begin{figure}[ht]
  \centering
    \includegraphics[width=3.5in]{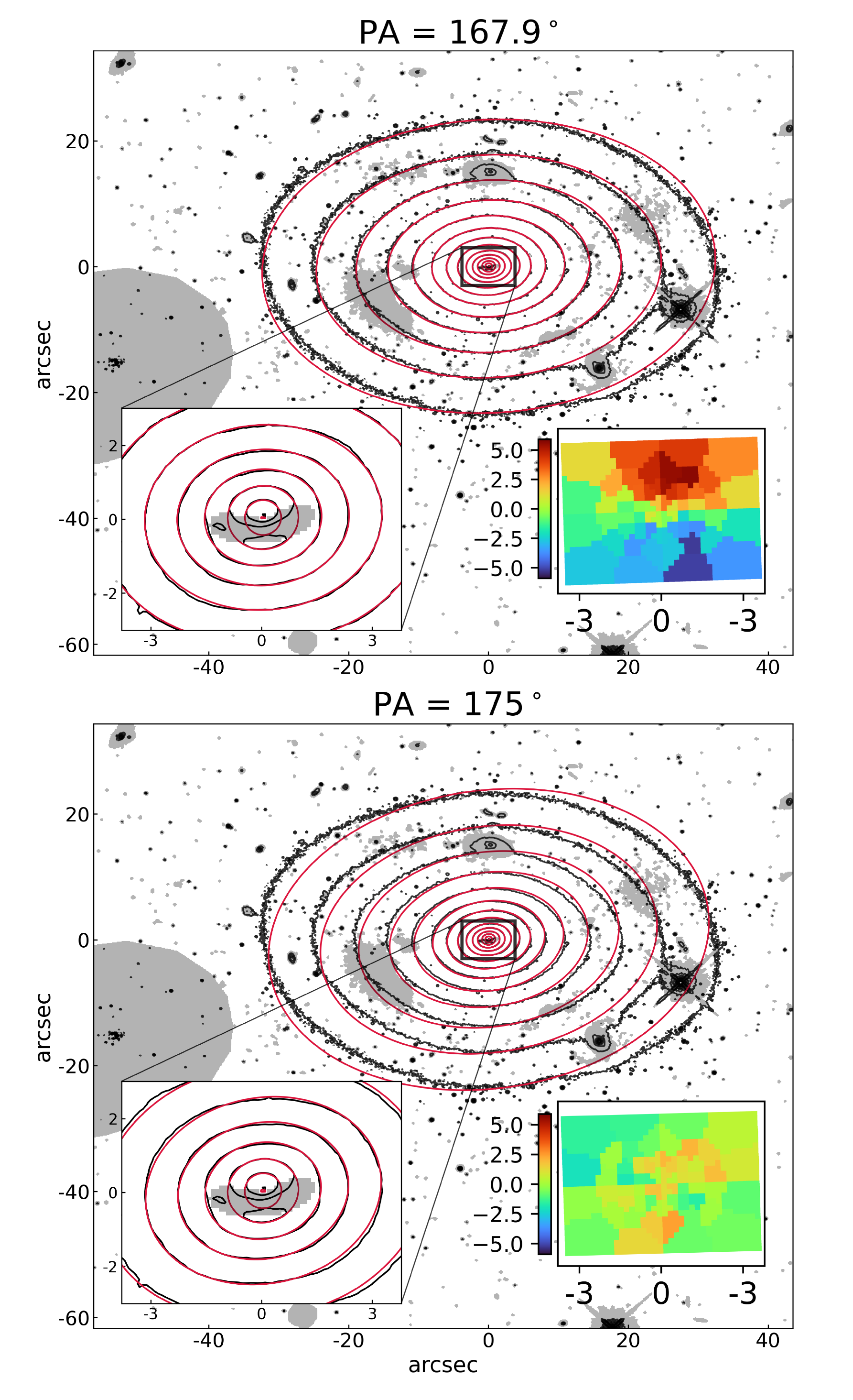}
    \caption{
    \cpm{Illustration of the non-alignment between the photometric PA and kinematic features.
    The upper panel repeats the \textit{HST} (black) and MGE model isophotes (red) in Figure~\ref{fig:MGE} and the non-bisymmetric component of the GMOS data in Figure~\ref{fig:bisymmetric_decomposition}. The best-fit MGE PA of the photometric major axis is $167.9^\circ$ in this fiducial case.  In the lower panel,
    we inflate the MGE PA to $175^\circ$ and plot the resulting model fits and non-bisymmetric map assuiming this PA.
    While the non-bisymmetric velocity pattern is minimized, this inflated PA provides a poor fit to the observed surface brightness profile.}
  } \label{fig:PA_rotation}
\end{figure}

\cpm{We have run an additional test to verify that the non-bisymmetric component of the kinematics show in Figure~\ref{fig:bisymmetric_decomposition} is due to a physical non-alignment between the photometric and kinematic axes and can not be simply ``rotated away".  
In this test,
instead of using the best-fit photomertic PA $167.9^\circ$ given by the MGE (see Sec.~2.1), we inflate the photometric PA to be $175^\circ$, a value that would minimize the magnitude of the non-bisymmetric component.
Using this inflated PA, we then refit the MGE and re-compute the non-bisymmetric component of our input kinematics.
Figure~\ref{fig:PA_rotation} compares the result for this test (bottom panel) with
that of our fiducial PA (top panel).
While the non-bisymmetric feature is indeed much reduced for ${\rm PA}=175^\circ$, 
the MGE isophotes for this PA provide a noticeably worse fit to the observed surface brightness profile at both at large and small radii. 
}

\citet{Lipkaetal2021} recently argued that edge-on axisymmetric models have a larger model flexibility than face-on projections
and thus can fit observational data better, biasing the recovered inclination
towards $i \sim 90^\circ$.
The rationale is that in edge-on models, the prograde and retrograde orbits have opposite velocities along the line of sight and contribute uniquely to the model's LOSVDs, whereas in face-on models, the two sets of orbits have negligible line-of-sight velocities, making them virtually interchangeable and
effectively reducing the number of unique orbits used in superposition and minimization routines. They reported a $\Delta \chi^2 \sim 30$ bias, favoring edge-on inclinations.
We have performed a parameter search including inclination as a fourth model parameter, sampling 1000 values from $i=[68^\circ,89^\circ]$ in the hypercube. Our regression find a best fit value $i = 87.6^{\circ}{}^{+0.9^\circ}_{-1.8^\circ}$, with a $\Delta \chi^2$ between the lowest and highest inclinations of $\sim 25$, slightly smaller than that reported in \citet{Lipkaetal2021}.  Despite this preference for edge-on inclinations in the axisymmetric models, our best-fit \mbh\ and \ml\ barely change when we include inclination as a free parameter: $\mbh = \left(2.4\pm0.5\right)\times 10^{9}\ \msun$ and $\ml = (2.23 \pm 0.1) M_\odot / L_\odot$. These are both consistent within the confidence intervals of the $i=70^\circ$ results, so our results are robust to choice of inclination angle. \

\subsection{Jeans Anisotropic Models}

We further model the stellar kinematics of NGC 2693 as an axisymmetric system using Jeans anisotropic modeling (JAM; \citealt{Cappellari_2008, Cappellari_2020}). JAM solves the Jeans equations assuming a velocity ellipsoid that is aligned with a cylindrical coordinate system ($R$, $z$, $\phi$) \cpm{or a spherical coordinate system ($R$, $\theta$, $\phi$). We adopt a cylindrically aligned velocity ellipsoid, which} is flattened along the $z$-axis and is characterized by the anisotropy parameter $\beta_z = 1 - (\sigma_z/\sigma_R)^2$, where $\sigma_z$ and $\sigma_R$ are the velocity dispersions parallel to the rotation axis and in the radial direction. JAM has the advantage of being computationally inexpensive and previous studies generally have found similar results between (axisymmetric) Schwarzschild models and JAM (e.g., \citealt{Seth_2014, Krajnovicetal2018a, Thater_2019}).

\begin{figure}
    \centering
    \includegraphics[width=3.3in]{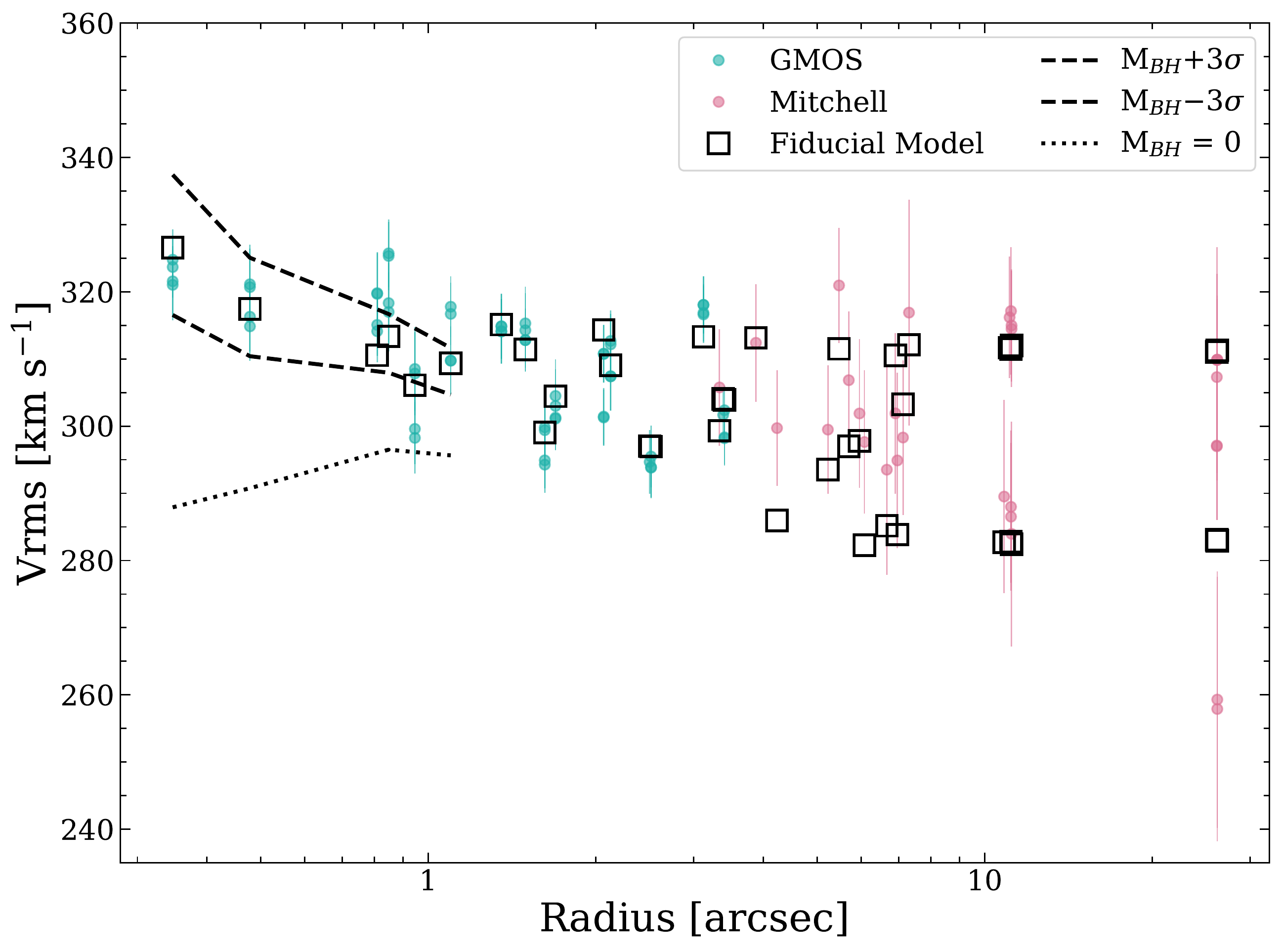}
    \caption{\cpm{Line-of-sight rms velocity ($V_{rms}$) determined from the GMOS (blue dots) and Mitchell (pink dots) IFS kinematics. The best-fit JAM model is shown with black open squares. Also shown are three JAM models with $\mbh$\ fixed to $0 \msun$ (dotted line), $2.1\times10^9\ \msun$ (the 3$\sigma$ lower bound; dashed line), and $3.7\times10^9\ \msun$ (the 3$\sigma$ upper bound; dashed line). The three models extend over all radii, although only the model predictions within the central region are plotted. Beyond $\sim 1\arcsec$, it is difficult to distinguish between the models. The best-fit model is a good match to the observations and the model without a SMBH underestimates $V_{rms}$ at the nucleus.}}
    \label{fig:JAM_vrms}
\end{figure}

In our model, the gravitational potential comes from the BH, stars, and dark matter. The galaxy's surface brightness from Table~\ref{tab:mge} is deprojected into a 3D stellar mass density given an inclination angle, $i = 70^\circ$, and $M^*/L$, while the dark matter halo is parametrized by the logarithmic profile in Equation~(\ref{eq:loghalo}). Thus, the free parameters in our model are: \mbh, \ml, $M_{15}$ (the dark matter mass enclosed within 15 kpc), and $\beta_z$. Given these parameters and the GMOS PSF, JAM predicts the second moment, which we compare to the observed $V_\mathrm{rms}$, with $V_\mathrm{rms} = \sqrt{(V^2 + \sigma^2)}$. We use the point-symmetrized $V$ and $\sigma$ from the GMOS and Mitchell observations, excluding the innermost Mitchell kinematics that spatially overlap with the GMOS kinematics. We additionally exclude the outermost four Mitchell bins as was done in the Schwarzschild models in earlier sections. 

The model parameters are optimized using Bayesian inference and the nested sampling code {\tt Dynesty} \citep{Speagle_2020}, which estimates posteriors and evidences. We adopted a likelihood $L \propto \exp(-\chi^2/2)$ where $\chi^2 = \Sigma_i (D_i - M_i)^2/\sigma_i^2$ and $D_i$ and $M_i$ are the observed and model $V_\mathrm{rms}$, respectively, and $\sigma_i$ is the $V_\mathrm{rms}$ uncertainty for each spatial bin. When running with {\tt Dynesty}, we use 500 live points and stop the initial sampling stage once reaching a threshold of 0.05, which is the log-ratio between the current estimated Bayesian evidence and the remaining evidence. The batch sampling stage is stopped when the fractional error on the posterior reaches 0.02. We assume uniform priors, with all free parameters sampled linearly. The best-fit values and 1$\sigma$ uncertainties are taken to be the median and 68\% confidence intervals of the posterior distributions, respectively. 

The results are shown in Table~\ref{tab:triax_results} \cpm{and the comparison between the best-fit model and the observed $V_\mathrm{rms}$ is given in Figure~\ref{fig:JAM_vrms}.} The model reproduces the data well, with a reduced $\chi^2$ of 1.07. \cpm{Figure~\ref{fig:JAM_vrms} also displays three models with \mbh\ set to 0 \msun, $2.1\times10^9$ \msun\ (the 3$\sigma$ lower bound), and $3.7\times10^9$ \msun\ (the 3$\sigma$ upper bound), with $M^*/L$, $M_{15}$, and $\beta_z$ fixed to the values from the best-fit JAM model in Table~\ref{tab:triax_results}. The \mbh\ $=0$ \msun\ case fails to match the kinematics in the inner region and further demonstrates the need for a black hole in the galaxy potential.} The \mbh\ and \ml\ from \cpm{the best-fit} JAM \cpm{model} are consistent within 1$\sigma$ of the axisymmetric Schwarzschild model results in Section~\ref{subsec:axi_orbit_modeling}, and $\beta_z$ falls within the range of values extracted from the best-fit axisymmetric Schwarzschild model. 
The $M_{15}$ value from JAM is lower than the Schwarzschild model result, but it remains consistent at the 2$\sigma$ level. \citet{Liepoldetal2020} found a similar result in the analysis of NGC~1453, with JAM favoring a value of $M_{15}$ half that inferred from the axisymmetric Schwarzschild models. 

We complete additional JAM runs to test assumptions made during the modeling. In our fiducial model, we fix $i = 70^\circ$, which is the inclination angle inferred from the dust disk, but we also test allowing $i$ to be a free parameter. We find a preference for $i = 81^\circ$, however all the angles for which the MGE could be deprojected fall within the 3$\sigma$ uncertainties. In addition, we test using a spatially varying anisotropy, with a parameter ($\beta_{z\mathrm{,in}}$) assigned to the MGE components with $\sigma_k^\prime < 3$\arcsec\ and another parameter ($\beta_{z\mathrm{,out}}$) attributed to the remaining MGE components. We then examine a case where $\beta_{z\mathrm{,in}}$ corresponded to the MGE components with $\sigma_k^\prime < 6$\arcsec. These choices were motivated by the previously run Schwarzschild models, which suggested a change in the anisotropy between a radius of $\sim 3-6$\arcsec. \cpm{Next, we fit to only the GMOS data, which extend to a radius of 3\farcs4, and we assume a spatially constant anisotropy.} Finally, we test including the Mitchell kinematics from the outer spatial bins during the fit, adopting an modified MGE constructed using a dust mask with fewer central pixels flagged, and increasing the number of live points and applying different sampling thresholds in {\tt Dynesty}. Even when changing the model in these various ways, we \cpm{nearly} always find consistent results at the 1$\sigma$ level with the fiducial model. \cpm{The exception is when we fit to only the GMOS data with a spatially constant $\beta_z$; we find that $M^*/L$ is consistent within the 2$\sigma$ uncertainties while the remaining parameters are in agreement at the 1$\sigma$ level with the fiducial model.}

\cpm{The above work assumed a cylindrically aligned velocity ellipsoid, but we also examine using spherically aligned JAM. In this case, we find a large anisotropy, with $\beta = 1-(\sigma_\theta/\sigma_r)^2 = 0.39$ and an order-of-magnitude smaller \mbh\ with 3$\sigma$ uncertainties that extend to 0 \msun. We also fit spherically aligned JAM to only the GMOS kinematics and recover the same results. When repeating the run and fixing $\beta = 0.0$, we find that \mbh\ is constrained with a best-fit value of $3.4\times10^9\ \msun$. In this case, the \mbh\ and remaining parameters are consistent with the fiducial (cylindrically aligned) JAM model given the 1$\sigma$ uncertainties.}

Despite the assumptions of cylindrically aligned JAM, the inferred \mbh\ and \ml\ match (at the 1$\sigma$ level) the results from the more complex axisymmetric orbit model in Section~\ref{subsec:axi_orbit_modeling} (Table~\ref{tab:triax_results}). The enclosed dark matter mass from JAM is $\sim40$\% lower than that from the axisymmetric orbit model, but it is within 2$\sigma$ uncertainties of the orbit model. 
As Table~\ref{tab:triax_results} shows, the uncertainties in the best-fit parameters from JAM tend to be much smaller than those from the axisymmetric orbit model.
We continue to see a shift in the \mbh\ compared to the best-fit value from the triaxial Schwarzschild model, with the JAM value being $75\%$ more massive than the value predicted from the triaxial modeling; see Section~6.2 for further discussion.

\section{Discussion}
\label{sec:six}

\subsection{Black Hole Scaling Relations}

To place the NGC~2693 SMBH on the $\mbh-\sigma$ relation, we use the luminosity-weighted velocity dispersion within $R_e$, $\sigma = 296 \kms$, from \citet{Vealeetal2017a} for NGC~2693.  This measurement was obtained from the same Mitchell IFS data used in this paper. 
The mass of the NGC~2693 SMBH is within 15\% of the value predicted by the mean $\mbh-\sigma$ relation in \citet{mcconnellma2013} and $\sim 5$\% above the relation in \citet{Sagliaetal2016}; it is within the intrinsic scatter of both relations, with values of 0.38 dex and 0.417 dex, respectively.

For the $\mbh-M_\text{bulge}$ relation, 
we use the total stellar mass of NGC~2693 from our best-fit triaxial model, $M^* = 7.2 \times 10^{11}\ M_\odot$, as the bulge mass.\footnote{This is $15\%$ larger than the stellar mass estimated from the ATLAS$^\text{3D}$ $M_K-$stellar mass relation \citep{cappellarietal2013}, using a $K$-band absolute magnitude of $M_K = -25.76$ \citep{Maetal2014}.}
The NGC~2693 \mbh\ is $\sim 25$\% smaller than the value predicted by the mean $\mbh-M_\text{bulge}$ relation of \citet{mcconnellma2013} and $\sim 18$\% smaller than the value predicted by the \citet{Sagliaetal2016} relation.  Again, this SMBH is within the intrinsic scatter of both relations, with values of $0.34$ and $0.535$ dex, respectively.

\subsection{Comparison of Triaxial and Axisymmetric Models}
\label{subsec:comparison}

There are few studies that compare \mbh\ determination from fully triaxial stellar dynamical models to axisymmetric models of the same galaxy.
The best-fit $\mbh$ for both M32 \citep{vandenboschdezeeuw2010} and NGC~1453 \citep{Liepoldetal2020,Quennevilleetal2021b} were unchanged when relaxing the assumption of axisymmetry, whereas \mbh\ in NGC~3379 increased by a factor of $\sim2$ in the triaxial case \citep{vandenboschdezeeuw2010}.  We note that the mass modeling performed for M32 and NGC~3379 did not simultaneously model the dark matter halo of the two galaxies.
In comparison, we make no assumptions on the dark matter halo of NGC~2693, and instead constrain the dark matter mass at $15$ kpc directly as was done for NGC~1453 \citep{Quennevilleetal2021b}.
Furthermore, the triaxial code of \citet{vandenBoschetal2008} had an incorrect scheme for mirroring orbits, which we fixed in the TriOS code used for NGC~1453 and NGC~2693 here.

In the case of NGC~3998,  \citet{Walshetal2012}
applied the triaxial code of \citet{vandenBoschetal2008} and considered different dark matter halos.  The grid-based parameter search did not allow for simultaneously varying all parameters in their model. While NGC~3998 was not modelled in the axisymmetric regime, the gas-dynamical measurement of \mbh\ disagreed with the stellar-dynamical value by a factor of $\sim4$ \citep{deFrancesco2006}. 

Recently, \citet{denBrok2021} applied the \citet{vandenBoschetal2008} code to the brightest cluster galaxy PGC~046832 to determine its intrinsic shape, central black hole mass, and orbital composition. The galaxy has a unique velocity map, exhibiting both a kinematically decoupled core and dramatic twists in the velocity field, suggesting a non-axisymmetric intrinsic shape. Their triaxial models prefer prolate galaxy shapes in the inner 10 arcseconds of the galaxy, and oblate shapes beyond 10 arcseconds, though these models only provide an upper bound on the black hole mass of $\mbh \lesssim 2\times 10^{9} M_\odot$. 
\cpm{While this disagrees considerably with the results from their best-fit axisymmetric models, which prefer $\mbh \sim 6\times 10^{9} M_\odot$, it remains to be seen if their triaxial result would change after the incorrect orbit mirroring in the \citet{vandenBoschetal2008} code 
and other issues discussed in \citet{Quennevilleetal2021b} are fixed.}

\cpm{In the case of NGC~2693, the best-fit orbit model in the axisymmetric limit 
and the best-fit JAM model
favor \mbh\ that is 40\%-70\% higher than the triaxial orbit model,
but the difference is within $\sim 2\sigma$ confidence level (see Table~2).  Similar comparison studies are needed from more galaxies to assess whether 
any systematic difference exists in \mbh\ values determined from different methods.}

\section{Summary}
\label{sec:seven}

We have reported detection of a SMBH with $\mbh = \left(1.7\pm0.4\right) \times 10^{9}\ \msun$ at the center of the massive, fast-rotating galaxy NGC~2693 targeted by the MASSIVE survey.
Using HST stellar light profiles and extensive IFS kinematic data covering a FOV from $\sim 150$ pc to 15 kpc as constraints (Section~\ref{sec:two}), we have performed triaxial orbit modeling with the \textit{TriOS} code to determine the galaxy's
internal stellar orbit structure, \mbh, \ml, dark matter content, and intrinsic 3D shape (Section~\ref{sec:three}). 
We modeled the gravitational potential of NGC~2693 with 6 parameters and performed a 6D Bayesian search using Latin hypercube sampling of $\sim 10,000$ galaxy models to find the model that best matches our input data (Section~\ref{sec:four}).

Despite NGC~2693 exhibiting properties typically indicating an intrinsic axisymmetric shape, we find 
the best-fit model to be triaxial with $T = 0.39 \pm 0.04$ and intrinsic axis ratios $p = b/a = 0.902\pm0.009$ and $q = c/a = 0.721^{+0.011}_{-0.010}$.
We find that triaxial models are needed to account for non-axisymmetric features seen in the residuals of our accompanying axisymmetric models (Figure~\ref{fig:bisymmetric_decomposition}). 
When limiting ourselves to axisymmetry, we find  $40\%$ larger best-fit black hole mass of $\mbh = \left(2.4 \pm 0.6\right) \times 10^{9}\ \msun$ from axisymmetric orbit modeling, and $75\%$ larger best-fit black hole mass of $\mbh = \left(2.9\pm 0.3\right) \times 10^{9} \msun$ from JAM modeling (Section~5); \cpm{both values are within $\sim 2\sigma$ confidence level of \mbh\ determined from triaxial modeling (Table~2)}.

\cpm{We have examined orbit flexibility in our galaxy models to assess possible effect of ``generalized degrees of freedom'' \citep{Ye1998,Spiegelhalter2002} on parameter determinations. Using a similar measure as \citet{Lipkaetal2021} to estimate the effective number of parameters, we find that our models in the axisymmetric limit have a similar behavior as \citet{Lipkaetal2021}, in which edge-on orientations tend to have higher model flexibility (Section~5.1).
Such varying model flexibility can be attributed to varying degeneracy between prograde and retrograde short-axis loop orbits as the line-of-sight approaches the symmetry axis. 
For triaxial models, however, we find 
the model flexibility to vary much less in the region around the best-fit models,
and our best-fit triaxial shape parameters change by less than $1\sigma$ in a number of preliminary tests. 
It is possible that the additional presence of \clt{box and} long-axis tube orbits in triaxial potentials has led to a weaker dependence of model flexibility on viewing angles.
We will report the full results in a subsequent paper.
}

This paper adds to only a handful of other stellar dynamical modeling studies not limited to axisymmetric galaxy shapes (Section~\ref{subsec:comparison}). Most of the remaining galaxies in the MASSIVE survey exhibit more prominent kinematic and photometric twists and less rotation compared to NGC~2693,
further providing evidence that massive early-type galaxies have triaxial intrinsic shapes.
More stellar dynamical measurements beyond the axisymmetric limit will inform whether the systematic differences in \mbh\ seen for NGC~2693 in this paper is a common occurrence.

\acknowledgments

J.D.P, C.M.L., C.-P.M., and M.E.Q. acknowledge support from NSF AST-1817100, HST GO-15265, HST AR-14573, the Heising-Simons Foundation, and the Miller Institute for Basic Research in Science. 
J.L.W. and S.C.D.A were supported in part by NSF grant AST-1814799.  M.E.Q. acknowledges the support of the Natural Sciences and Engineering Research Council of Canada (NSERC), PGSD3-517040-2018. This work used the Extreme Science and Engineering Discovery Environment (XSEDE) at the San Diego Supercomputing Center through allocation AST180041, which is supported by NSF grant ACI-1548562.  Portions of this research were conducted with the advanced computing
resources provided by Texas A\&M High Performance Research Computing.
This work is based in part on data obtained at the international Gemini Observatory, a program of NSF’s NOIRLab, which is managed by the Association of Universities for Research in Astronomy (AURA) under a cooperative agreement with the National Science Foundation on behalf of the Gemini partnership: the National Science Foundation (United States), the National Research Council (Canada), Agencia Nacional de Investigación y Desarrollo (Chile), Ministerio de Ciencia, Tecnología e Innovación (Argentina), Ministério da Ciência, Tecnologia, Inovações e Comunicações (Brazil), and Korea Astronomy and Space Science Institute (Republic of Korea).  This work is based in part on observations made with the NASA/ESA Hubble Space Telescope, obtained at the Space Telescope Science Institute, which is operated by the Association of Universities for Research in Astronomy, Inc., under NASA contract NAS5-26555. These observations are associated with program GO-14219.

\software{Astropy \citep{astropy:2013, astropy:2018},
Dynesty \citep{Speagle_2020},
Galfit \citep{peng02},
jampy \citep{cappellari2008},
Matplotlib \citep{Hunter:2007},
mgefit \citep{Cappellari2002},
NumPy \citep{harris2020array}.}

\bibliography{N2693}

\end{document}